\documentclass[12pt]{article}
\usepackage[latin1]{inputenc}
\usepackage{amsmath}
\usepackage{amsfonts}
\usepackage{amssymb}
\usepackage{graphicx}
\usepackage{geometry}
\usepackage{hyperref}
\usepackage[english]{babel}
\usepackage[bottom]{footmisc}
\title{Categorical Exploratory Data Analysis: From Multiclass Classification and Response Manifold Analytics perspectives of baseball pitching dynamics.}
\author{Fushing  Hsieh\footnotemark[1], Elizabeth P. Chou\footnotemark[2]}
\date{}
\begin{document}
\maketitle

\begin{abstract}
From two coupled Multiclass Classification (MCC) and Response Manifold Analytics (RMA) perspectives, we develop Categorical Exploratory Data Analysis (CEDA) on PITCHf/x database for information content of Major League Baseball's (MLB) pitching dynamics. MCC and RMA information contents are represented by one collection of multi-scales pattern categories from mixing geometries and one collection of global-to-local geometric localities from response-covariate manifolds, respectively. These collectives shed lights on the pitching dynamics, and maps out uncertainty of popular machine learning approaches. On MCC setting, an indirect-distance-measure based label embedding tree leads to discover asymmetry of mixing geometries among labels' point-clouds. A selected chain of complementary covariate feature groups collectively brings out multi-order mixing geometric pattern categories. Such categories then reveal true nature of MCC predictive inferences. On RMA setting, multiple response features couple with multiple major covariate features to demonstrate physical principles bearing manifolds with a lattice of natural localities. With minor features' heterogeneous effects being locally identified, such localities jointly weave their focal characteristics into system understanding, and provide a platform for RMA predictive inferences. Our CEDA works for universal data types, adopts non-linear associations and facilitates efficient feature-selections and inferences.
\end{abstract}

\footnotetext[1]{Correspondence: Department of Statistics, University of California at Davis, CA, 95616. E-mail: fhsieh@ucdavis.edu}.
\footnotetext[2]{Department of Statistics, National Chengchi University, Taiwan}

\section{Introduction}
The author of well-known 1977 book: Exploratory Data Analysis (EDA) \cite{tukey77}, John W. Tukey in his 1962 paper \cite{tukey62}: The future of Data Analysis, discussed and made clear-cut separations among mathematics, mathematical statistics and data analysis. He claimed that Data Analysis, as his central interest, is indeed a science. He further wrote that:

``Data analysis is a larger and more varied field than inference, or incisive procedures, or allocation."

This statement rings much louder now than ever before in the current era of Data Science. He also reminded readers that data analysts must put ``reliance upon the test of experience as the ultimate standard of validity''. This statement paves the foundation for Data Science and explains why Data Analysis is independent of mathematics.

However, 60 years into data analysis after his paper, we neither have well-developed resolutions for major problems listed in his paper: spotty data in more complex situations and multiple response data, nor unified fundamental concepts and computational paradigms as principles to sustain Data Analysis or Data Science as a stand-alone discipline of science. One encouraging sign is that data analysis has been widely permeating and drastically expanding in all sciences. Nevertheless, within each real-world data analysis, we see that the danger sign is constantly blinking and signaling: Data Analysis is a still underdeveloped field.

One direction to view such underdevelopments of data analysis is to try to analyze a real database from a complex physical system. Within such a system, physics helps us understand what are underlying principles and how they work. However, we often find that we miss how these principles couple and link together in working out details. The phrase of ``The devil is in the detail'' seems to vividly capture what is exactly missing. Can data analysis discover such principles and missing details of their linkages? Would data analysis be capable of aiming for full understanding of such a system as one whole? We attempt positive answers to both questions through a real complex system in this paper. Also we propose and demonstrate an universal foundation for computing and constructing these answers.

The real example system considered here is the Major League Baseball's (MLB) pitching dynamics in USA. The MLB has been recording and storing every single pitch in its 30 stadiums into its public available PITCHf/x database since year 2006. From physics, Newtonian laws of forces are known as the fundamental principles of pitching dynamics. But, we want explicitly see how Magnus force of spin \cite{briggs59}, which is induced from these principles, govern a baseball trajectory. That is, we want well-patched details of pitching dynamics contained in PITCHf/x database without invoking an existing set of differential equations from physics and aerodynamics literatures. Such computable details would make us see which factors make a pitch move and curve the way it does, and understand why some characteristic features are pitcher idiosyncratic. Here well-patched details mean multiscale and global-to-local pattern information that shed lights on how principles work out in concert through data analysis. We collectively term such details: ``information content'' contained in data.

In line with Tukey's statement, as would be illustrated in the next section, PITCHf/x's information content regarding pitching dynamics contains at least two major perspectives: 1) what factors can efficiently characterize a pitcher's pitches, and 2) how actually the underlying physical principles work out? The first perspective falls into one major machine learning topic, called multi-class classification (MCC). The second perspective can be prescribed as about how a group of possibly highly associative response features or variables links to another group of covariate features. This perspective, which is termed Response Manifold Analytics (RMA), is not yet well developed in literature.

Here response-vs-covariate features are determined by clear spatial-temporal separations. That is, covariate features are measured at pitcher's mound, while response features are measured at the home plate, where a catcher receives a pitched baseball. With such temporal and spatial separations, RMA is exactly one multiple response problem. In fact, data from a complex system is typically embedded with a multiple response problem. Conversely, a multiple response problem is typical related to a complex system, in which intertwined physical or mechanical linkages are expected. Hence, presences of principle-bearing geometric manifolds, as would be clearly seen below, are expected to be involving in RMA's information content.

Though, information content contained in data derived from a complex system is in general not mathematically defined, its merits go far beyond statistical inferences and machine learning based predictions. Its visible and explainable graphic displays allow experts and amateurs to read out deterministic and stochastic structures contained in data. As a byproduct, such structural information explicitly exposes potential validity and inevitable violations of statistical modelings as well as black-boxed uncertainty of machine learning approaches. This concept of information content indeed coherently reflects the quoted statement on data analysis at the beginning of this section.

\section{Information content of PITCHf/x Complex system from perspectives of MCC and RMA}
There are 21 features measured in PITCHf/x database engineered toward pitching dynamics. These 21 features are either bio-mechanical or physical in nature. Here bio-mechanical features are those measured related to pitching gestures at the moment of pitching on the pitcher's mound, such as horizontal and vertical coordinates of a releasing point:\{``x0'', ``z0''\} ; $x-$, $y$- and $z-$ directional releasing speeds : \{``vx0'', ``vy0'', ``vz0''\}, and spin created by using fingers sliding against skin of baseball:\{``spin\textunderscore dir'', ``spin\textunderscore rate''\}, and several others. While physical features are referring to features that are measured at the home plate, like horizontal and vertical movements: \{``pfx\textunderscore x'', ``pfx\textunderscore z''\}; degree of curve in trajectory: \{``break\textunderscore length'', ``break\textunderscore angle''\}; accelerations: \{``ax'', ``az''\} and final speed \{``end\textunderscore speed''\}. It is evident that bio-mechanical and physical feature are temporally and spatially separated.

Even though pitching dynamics is universally governed by Newtonian physics, different pitch-types seemingly reveals their own physical and bio-mechanical characteristics and distinctions. Here we separate pitching dynamics with respect to three pitch-types: fastball, slider and curveball, in the order of decreasing speed. Their distinctions are intrinsic and beyond their speed differences. Such intrinsic distinctions are to be illustrated through graphic displays. Throughout this paper, computational results are all represented via graphic displays in order to reveal many visible aspects of information content pertaining to each pitch-type specific dynamics. Such displays, either in a form geometry or network, amount to confirm one vital and relevant concept: {\bf distinct feature sets manifest different perspectives of information}. By collecting and synthesizing all vital perspectives of pattern information, we hope to arrive at current experts' knowledge and understanding, or even go beyond.

We begin with building a road-map for our exploratory data analysis. This map is based on a $21\times 21$ mutual conditional entropy matrix of nonlinearity based associative measurements developed in \cite{hsiehshanyu}. As would be elaborated below, the lower mutual conditional entropy value between two features is, the higher associative they are. Such a matrix can be used as a ``distance'' matrix when applying the hierarchical clustering (HC) algorithm. The resultant hierarchical clustering tree indeed provides multiscale compositions of feature clustering or groupings, which bear clearer bio-mechanical and physical meaning than results derived from principle components based dimensional reduction approaches and factor analysis. This is one key of obtaining explainable and visible computational results. These multiscale clustering compositions have significantly reduced number of possible combinations of features collectively. Therefore, block-patterns found on a mutual conditional entropy matrix become a road map that make efficient explorations for information content possible.

For completeness, we briefly mention the idea and concept of entropy based association between two features or variables. Building a contingency table has been one simple classic way of exploring and exhibiting two variables' potential nonlinear associative relation, see \cite{hsiehshanyu}. This contingency table approach is fundamental in the sense that it works for all data types: continuous, discrete and categorical. When involving an 1-dim continuous variable, it needs to be first transformed into a discrete or categorical one. We strongly suggest to perform this transformation via its possibly gapped histogram, which brings out 1D data's hidden categorical structure through a proper piecewise linear approximation to its distribution function, see details in \cite{hsiehroy}.

Let categories belonging to the two variables be arranged on row-and column-axes of a contingency table, respectively. Given a row, the Shannon entropy of cell-proportions with respect to categories of the column-variable measures a degree of associative uncertainty conditioning on the row-specific category of row-variables. So the directed row-to-column associative measure is calculated as the weighted sums of row-specific Shannon entropies, and then re-scaled with respect to Shannon entropy of column variable, that is, the Shannon entropy of the column-proportions. The column-to-row association is likewise calculated via column-specific Shannon entropies.

By simply averaging the two directed entropies: row-to-column and column-to-row, we arrive at the so-called mutual conditional entropy (MCE) of these two variables. Thus, all these directed and mutual conditional entropies are invariant with respect to row-wise and column-wise permutations onto the contingency table. This is why such associative measures can accommodate all nonlinear associations when working for all data types, see \cite{hsiehshanyu} for details and their merits in inferences.

It is crucial to note here that {\bf such a direct association measure can be evaluated from a group of response feature to a group of covariate features} as well. Since a group of feature could be synthesized into a categorical variable by applying Hierarchical Clustering algorithm onto a corresponding data matrix, which is built by arranging all features on its column-axis, and all subjects' data-points on row-axis. The HC tree on the row-axis offers multiple versions of categorizing variables upon choices, see details in \cite{hsiehshanyu}.

\begin{figure*}[t]
		\centering
        \includegraphics[width=2in]{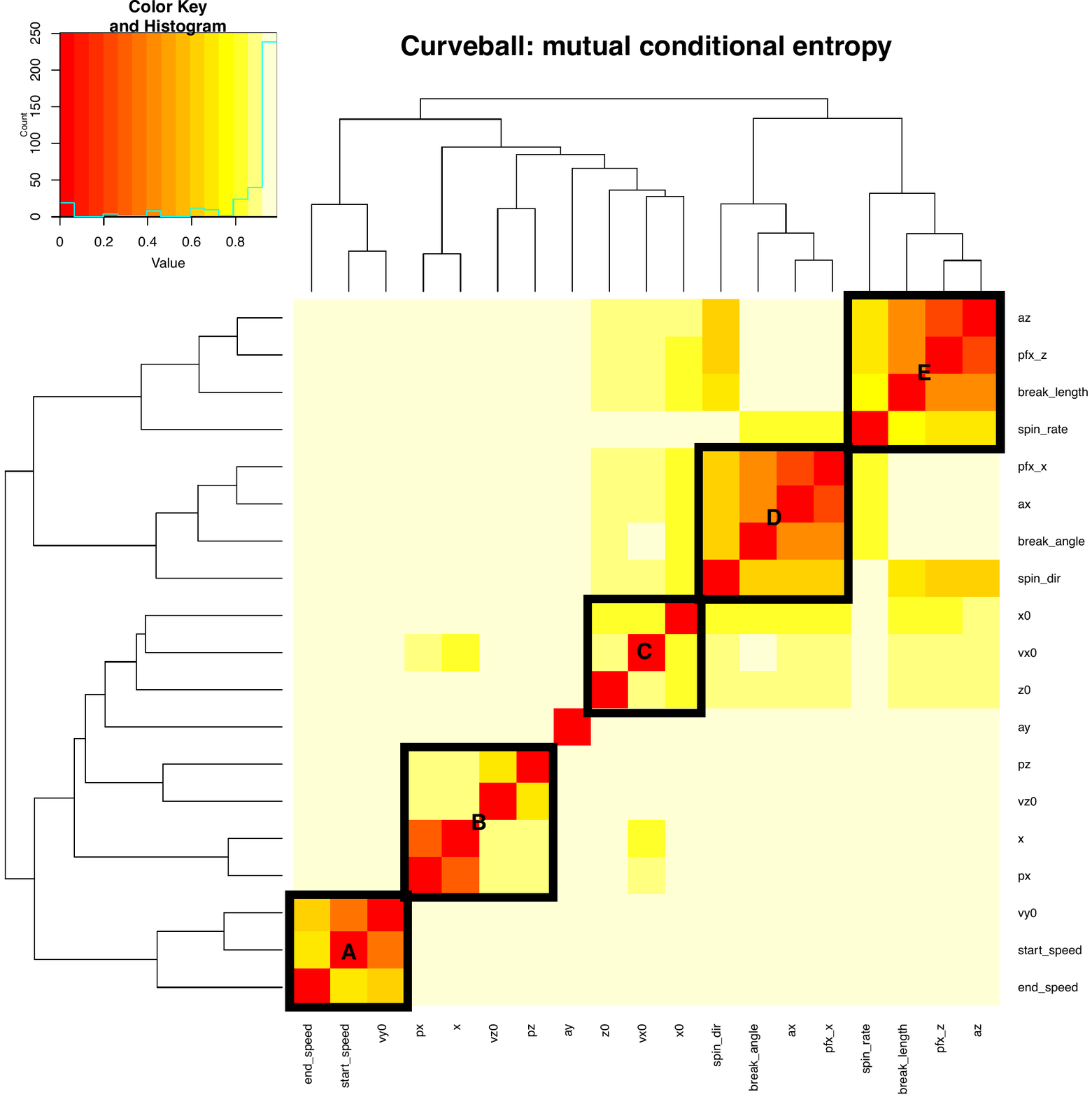}
		\includegraphics[width=2in]{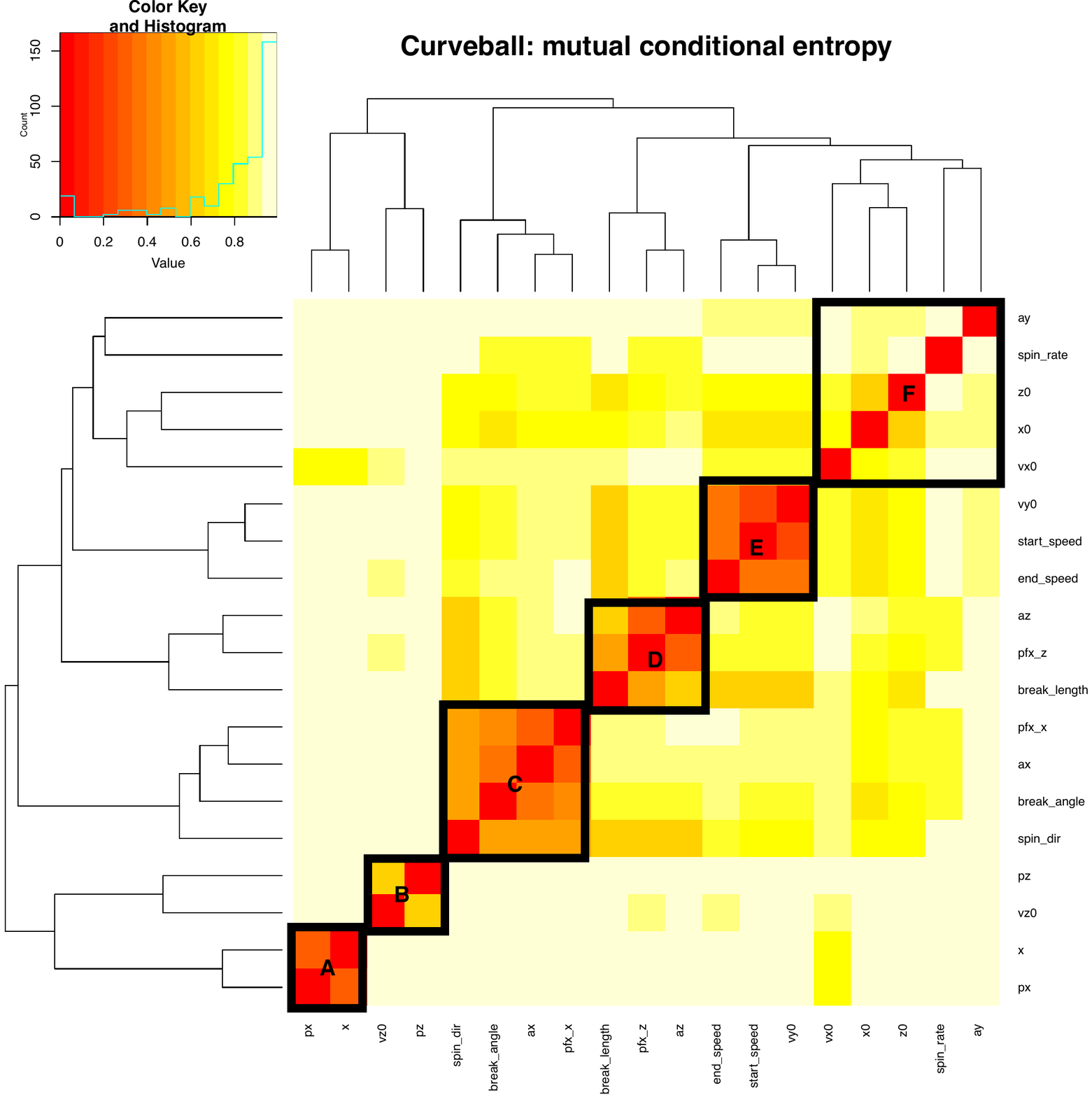}
        \includegraphics[width=2in]{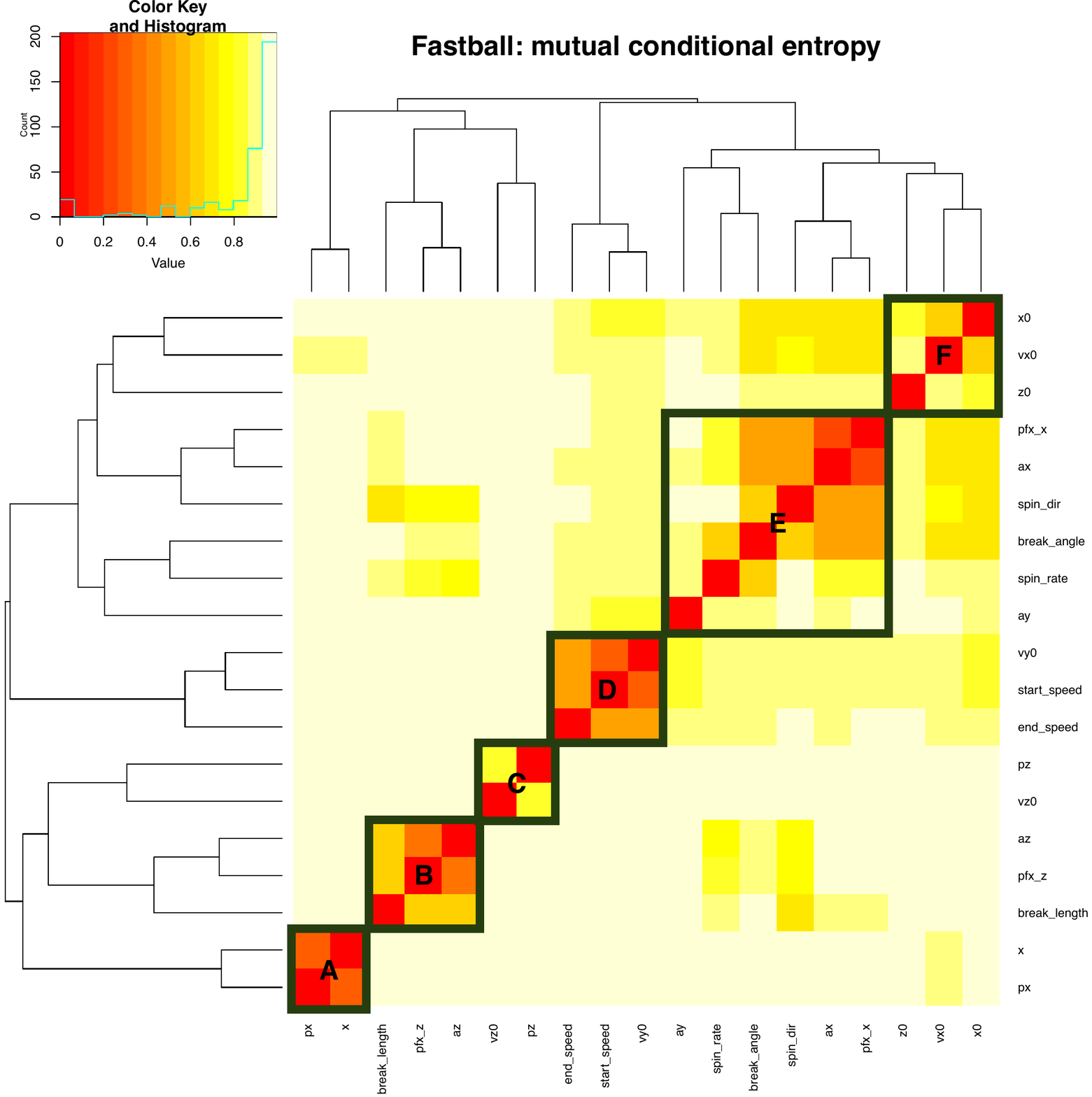}
		\caption{Three mutual conditional entropy matrices with marked synergistic feature-sets: (A) slider; (B) curveball; (C) fastball.}
		\label{fig:MCEM}
	\end{figure*}

As the lower mutual conditional entropy of a pair of features is, the higher association they have. Therefore, we can take such an entropy measure as one some sort of ``distance'' among all features. Hence, when the symmetric mutual conditional entropy matrix serves as a distance matrix for applying Hierarchical Clustering (HC) algorithm, we superimpose the resultant HC tree onto its row and column axes. Both trees are seen to frame and reveal multiscale block patterns upon the $21\times 21$ matrix lattice, see panels-A\& B \& C of Fig. \ref{fig:MCEM} on three pitch type: fastball, curveball and slider, respectively.

The three mutual conditional entropy matrices of Fig. \ref{fig:MCEM} are permuted and marked with block patterns indicating highly associative feature-groups among the 21 features. With very slightly distinctions across the three pitch-types, each block-based feature-groups should be intuitively taken as one constituting mechanism of pitching dynamics. For instance, the group \{``x0'', ``z0'', ``vx0''\} is a purely bio-mechanical across all pitch-types, while many blocks contain joint bio-mechanical and physical mechanisms by involving temporally and spatially separated features. This phenomenon is essential for exploring information content in PITCHf/x database. Below, we list feature-triplets accommodating such mechanisms, and then explicitly show these mechanisms via graphic displays:
\begin{description}
\item[a:][Bio-mechanics] ``z0'', ``x0'' and ``vx0'';
\item[b:][Bio-mechanics and laws of forces] ``x0'' and ``start\textunderscore speed'' toward ``end\textunderscore speed'';
\item[c:][Bio-mechanics and Laws of forces] ``spin\textunderscore rate'' and ``start\textunderscore speed'' toward ``end\textunderscore speed'';
\item[d:][Magnus effect] ``spin\textunderscore dir'' toward \{``pfx\textunderscore x'', ``pfx\textunderscore z''\}.
\end{description}

\begin{figure*}[h!]
\centering
\includegraphics[width=3.1in]{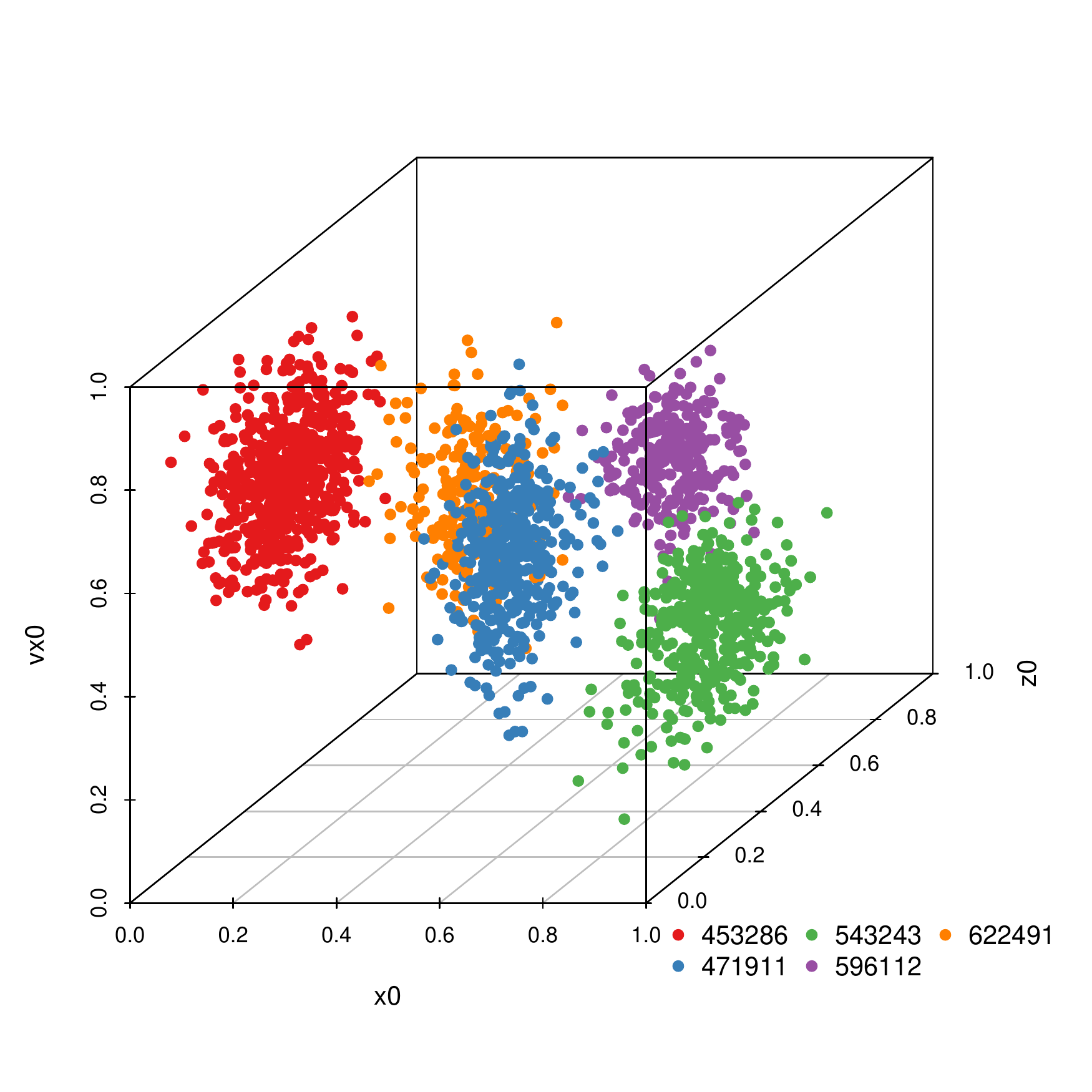}
\includegraphics[width=3.1in]{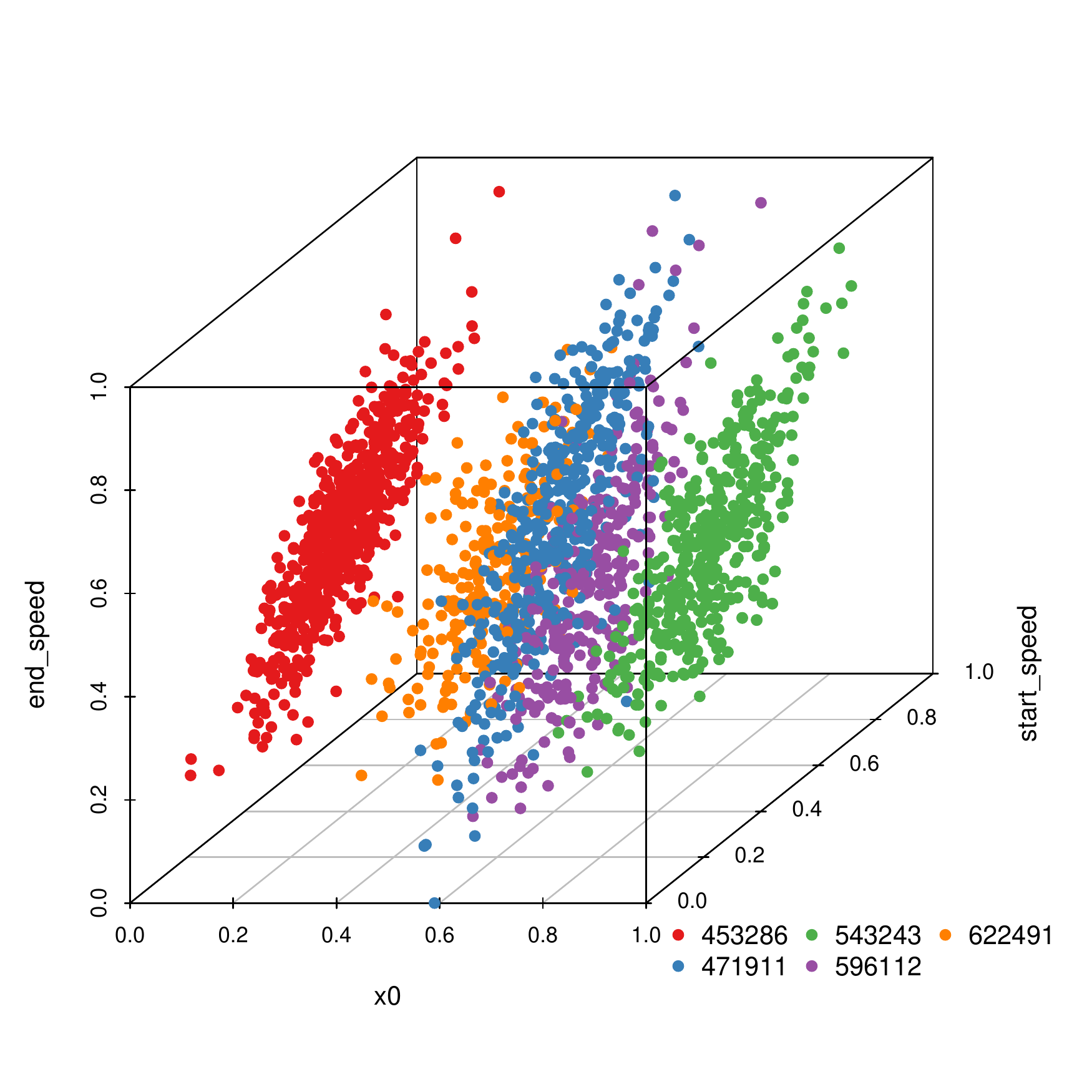}
\includegraphics[width=3.1in]{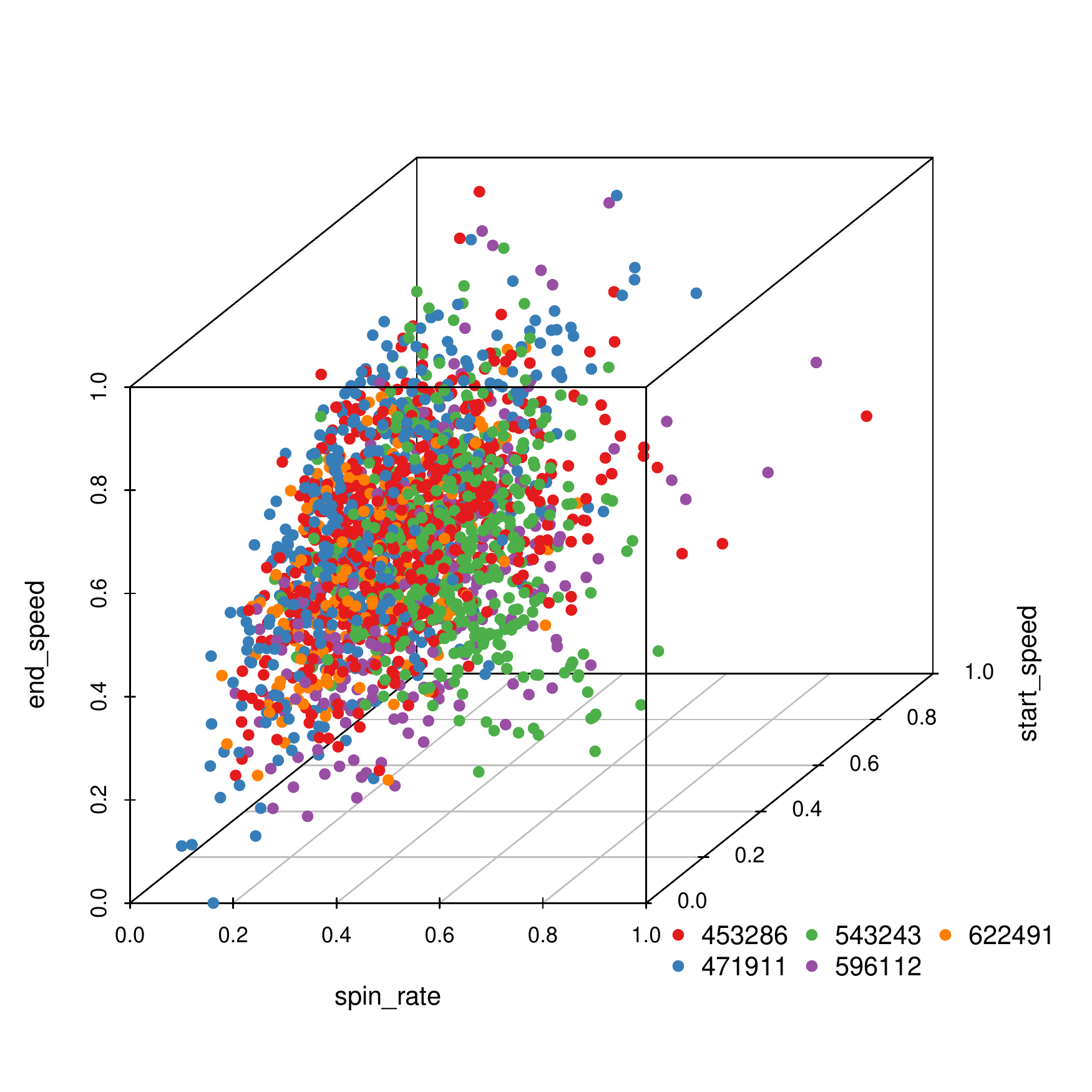}
\includegraphics[width=3.1in]{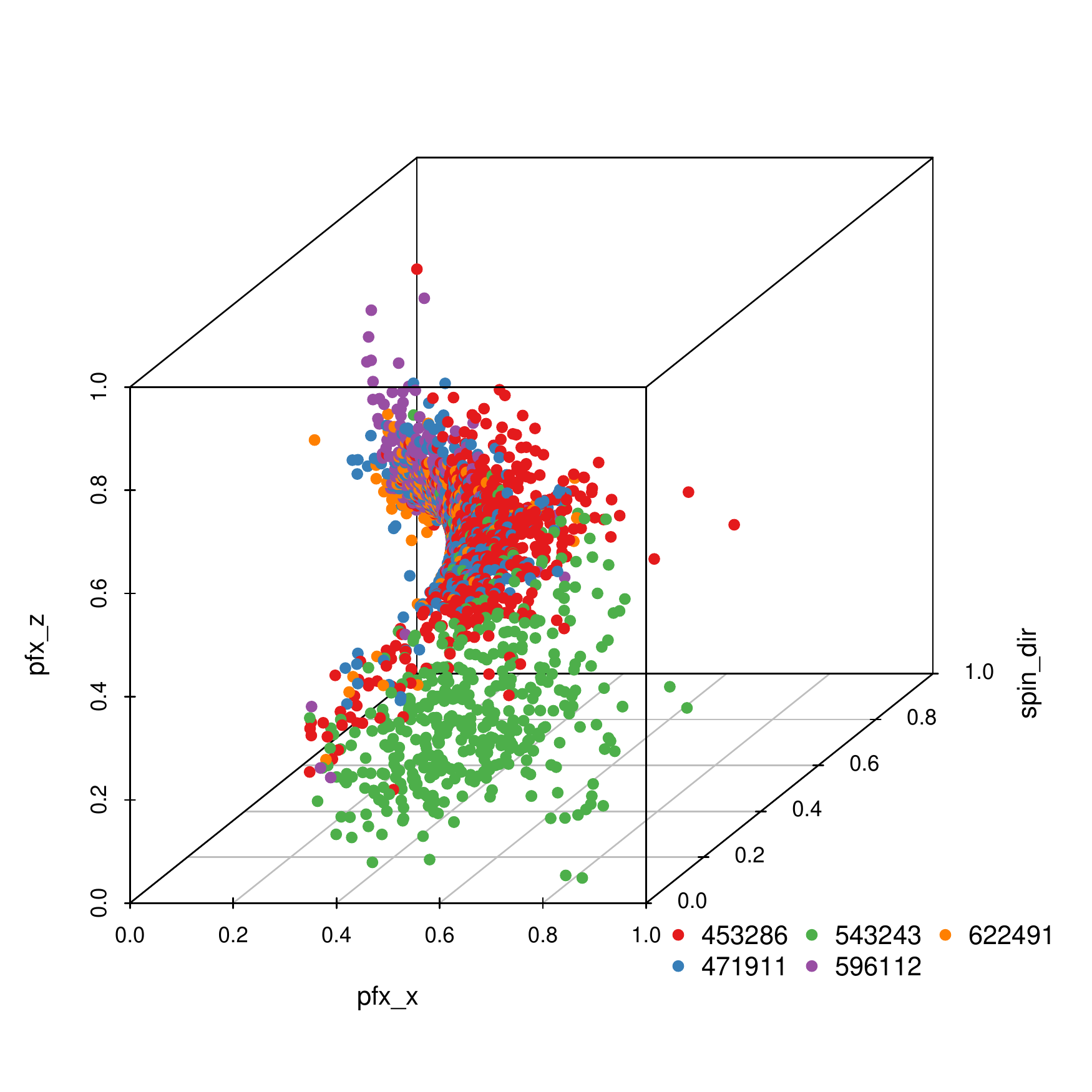}

\caption{Slider's four 3D geometries of five triplets of features (url-address for rotatable 3D plot):(A)\{``z0'', ``x0'', ``vx0''\}; (B)\{``x0'', ``start\textunderscore speed'', ``end\textunderscore speed''\}; (C)\{``spin\textunderscore rate'', ``start\textunderscore speed'', ``end\textunderscore speed''\}; (D)\{``pfx\textunderscore x'', ``spin\textunderscore dir'', ``pfx\textunderscore z''\}. See corresponding rotatable 3D plots in (\url{https://rpubs.com/CEDA/baseball}).}
\label{fig:infogeoS}
\end{figure*}

We illustrate aforementioned mechanisms of slider pitch-type. Slider in general gives rise to much more diverse and drastic horizontal movement than curveball and fastball do. This pitch-type has a wide range of ``spin\textunderscore dir'', while curveball has only top-spin (top-to-bottom wheeling in view of a catcher) and fastball has only back-spin (bottom-to-top wheeling). Data-points of feature-triples are color-coded with respect to pitcher-IDs, as shown in panels of Fig. \ref{fig:infogeoS} and corresponding rotatable 3D plots in (\url{https://rpubs.com/CEDA/baseball}).

These four geometries pertaining to the four feature-triplets in Fig. \ref{fig:infogeoS} and (\url{https://rpubs.com/CEDA/baseball}) are evidently distinct. One natural and essential question is: How do these geometries reveal information contents of individual mechanisms? To address such a question, we first briefly point out their visible geometric structures and potential information conveyed by the four panels of Fig. \ref{fig:infogeoS}, respectively.
\begin{description}
\item[Panel-A] The separating point-clouds in panel (A) indicates that the bio-mechanical oriented feature-set is informative for distinguishing among pitcher-IDs. It is logical and not surprising that features related to pitching gestures, like ``x0'' and ``z0'', are pitcher specific.

\item[Panel-B] The expected linear relation between ``start\textunderscore speed'' and ``end\textunderscore speed'' forms a shuttle shape via a cross section view. The interesting fact is that the geometry of shuttle composition is basically determined by pitcher-ID specific ``x0''. So this triplet of bio-mechanical and physical features also provides one perspective of information about pitcher-IDs. But it is not as informative as the triplet of features in panel-A.

\item[Panel-C] When replacing the ``x0'' with ``spin\textunderscore rate'', the geometry becomes drastically different. We see a 3D geometry with defusing colored dots (pitcher-IDs) all over. This fact indicates that this triplet of features is no longer informative for pitcher-IDs. Also this geometry simply demonstrates that ``spin\textunderscore rate'' even doesn't play any minor role whatsoever linking to the linearity from ``start\textunderscore speed'' toward ``end\textunderscore speed''.

\item[Panel-D] The sharp 2D spiral manifold in $R^3$ evidently implies that this triplet of features embraces one system of unknown mathematical equations that indeed depicts key physical mechanisms of pitching dynamics. Such a functional system might involve with features beyond \{``pfx\textunderscore x' and ``pfx\textunderscore z',``spin\textunderscore dir''\}. Since ``spin\textunderscore rate'' is also naturally expected to play a major role in pitching dynamics. How to explore major and minor candidate features involving in this complex system? Can we explicitly expose such involvements without mathematical formula?

\end{description}

From information content perspective, the panels-A\& B of Fig. \ref{fig:infogeoS} are seemingly informative under a MCC setting because of nearly separated point-clouds with respect to (color-coded) pitcher-IDs, while panels-B \&-D of Fig. \ref{fig:infogeoS} are essential for the response manifold analytics (RMA) setting because of both manifolds revealing evident geometric signatures. In particular, as shown via rotatable 3D plots in (\url{https://rpubs.com/CEDA/baseball}), the 3D plot of feature-triplet: \{``spin\textunderscore dir'', ``pfx\textunderscore x'', ``pfx\textunderscore z''\} is a sharp 2-dimensional manifold in $R^3$. This 2D manifold is indeed in accord with aerodynamic phenomenon first described by German experimental scientist H. G. Magnus in 1852. This so-called Magnus effect is a force generated by a spinning object (baseball) traveling through a viscus fluid (air). The force is perpendicular to the velocity vector of the object. So its spin direction on the 2D plane dictates a specific combination of vertical and horizontal movements on the traveling object. That is, in view of a catcher at the home plate, fastball seemingly goes up against earth's gravity, curveball seemingly has been pulled down harder than gravity could, while slider could go wild via a combination of evident vertical and horizontal movements. Other pitch-types, except the knuckleball,  such as change-up, sinker,...etc. are all characterized by various kinds of spin created by distinct ways of holding and releasing a baseball. The only pitch-type having zero spin is the knuckleball. By missing spins as stabilizers, air resistance makes knuckleball's trajectory very unpredictable, even for an experienced catcher.

Further, it becomes evident via the block-marked synergistic feature-groups in Fig. \ref{fig:MCEM} that the two triplets of features: \{``spin\textunderscore dir'', ``break\textunderscore length'',``break\textunderscore angle''\} and \{``spin\textunderscore dir'',``ax'',\\``az''\} also give rise to nearly 2D manifolds in $R^3$, respectively. Even though they are not as sharp as the manifold of \{``spin\textunderscore dir'',``pfx\textunderscore x'', ``pfx\textunderscore z''\}, these three manifolds have rather similar geometric structures. This observation is somehow surprising at first. Nonetheless, the underlying reason becomes intuitive from panels of Fig. \ref{fig:MCEM}. Since ``spin\textunderscore dir'', ``break\textunderscore angle'', ``ax'' and ``pfx\textunderscore x'' are in one synergistic feature group, while ``break\textunderscore length'', ``az'' and ``pfx\textunderscore z'' are in another group. It is worth noting that all these member features are commonly highly associative with ``spin\textunderscore dir'' and ``spin\textunderscore rate''. That is, these physical perspectives of understanding constitute information content and should be extracted from these manifolds via response manifold analytics (RMA). We are also intrigued by minor effects potentially exhibiting on localities of manifolds, such as pitcher-IDs as evidently seen in panel-D of Fig. \ref{fig:infogeoS}. Since heterogeneity is ubiquitous in any complex system \cite{anderson72}. These effects are also component-parts of physics of pitching dynamics. Thus, extracting information content in full from these manifolds under a MRA setting is a rather challenging issue.

We computationally resolve challenging issues of extracting information content under the RMA and MCC settings via an unified paradigm. This unified computing paradigm is called Categorical Exploratory Data Analysis (CEDA) because its basis primarily relies on computable categorical patterns contained in data. An example of categorical patterns is the contingency table used in the aforementioned mutual conditional entropy (MCE) matrix. Further, as would be seen below, MCC information content is represented via a collective of mixing geometric pattern categories, while RMA information content is a collective of pattern information extracted from a lattice of localities categorically framed upon manifolds.

\section{Multiclass Classification (MCC)}
A multiclass classification (MCC) setting can be found in the heart of majority of real world complex systems. Its label space, which contains a collection of labels, such as species in nature, diseases in medicine or pitchers in Major League Baseball (MLB), knowingly or unknowingly capsulizes many defining characteristics of its members through a labeling process. Two major types of labeling are reflected through two major data types: structured and unstructured, respectively. MCC's unstructured data can be an image from nature, subject or brain or issue scanning, or an article from newspaper, or website \cite{deng10}. Labeling on an image, article or website in general is subjective. Thus, MCC information content of an unstructured data set is likely arguable.
	
In sharp contrast, a MCC setting with structured data consists of $K$ pre-selected or carefully engineered features upon subjects respectively belonging to $L$ labels. So each data point under a label is a $K$-dim vector of measurements. The fundamental differences between unstructured and structured data rest on the following fact: The task of labeling of $L$ labels via selecting or engineering $K$ features is a way of knowingly reflecting and encapsulating curator's knowledge and intelligence upon a system of interest into data. For instance, an expert names a species, a medical doctor names a disease, even a pitcher pitches a baseball pitch. Ideally, their labeling constitute critical acts by nature or humans to capturing and encoding knowledge underlying natural systems or societal phenomena.
	
In this section, we want to computationally uncover and decode knowledge and intelligence underlying MCC settings with structured data. To a great extent, this is a way of understanding a labeling process of interest in a reverse engineering fashion. Such endeavors make Multiclass Classification (MCC) a major topic in the machine learning \cite{lecun98,Weiberger09,bengio10}. Recently, MCC has become a chief part of foundations for Artificial Intelligence (A.I.) \cite{russell09}. Though, MCC setting with structured data has been ardently studied \cite{cisse13,gupta14}, its information content has never been explicitly discussed and presented.
	
Here the MCC information content of the structured database PITCHf/x is needed in order to shed lights on questions like: How to precisely characterize chief distinctions among MLB pitchers' pitching dynamics? Can we make such characteristics and distinctions visible and readable? Is it possible to explain each of our predictive decision-makings with a pattern-based reason? Can we make no mistakes in identifying labels?

When $L$, $K$ and $N(=N_1+...+N_L$), the total number of data points, are not too small, such MCC information content indeed can be rather complex. One way of perceiving such complexity is through a mixing geometry of all labels' point-clouds determined by a subset of $k (\leq K)$ features. Such a mixing geometry surely exhibits potentially non-linear associative relations among the $k$ features because such associations shape each label's $k$-dim point-cloud individually and in turn dictate the mixing geometry among all point-clouds collectively.
	
Given that one feature-set offers one perspective of mixing geometry and there are $2^{21}-1 (>10^{6})$ possible feature-sets, so there are too many and too diverse candidate perspectives of MCC information content to be looked at and chosen from. Further, due to temporal and spatial separations, these non-linear feature-to-feature associations are likely directed, so their mixing geometric patterns indeed is likely asymmetric. That is, views of one single mixing geometry can vary from point-clouds' standpoints. As would be seen below, such hardly known mixing geometric asymmetry indeed help us to see through the complexity of MCC information content, only if we carefully choose which perspectives to look into.

The underlying reason is that such asymmetry consequentially gives rise to complementary feature-sets: one feature-set's mixing geometry shows clear separations among a sub-group of pitchers' point-clouds, and at the same time shows intense mixing among another subgroup of pitchers', while a complementary feature-set's mixing geometry reveals the opposite patterns. By identifying such pair of complementary feature-sets, we can arrange them into a chain, so that the uncertainty caused by mixing among point-clouds pertaining to the first feature-set can be significantly reduced and minimized due to different views provided by the latter feature-set, and so on. These are essential and brand new aspects of full MCC information content.
	
At this stage of knowledge, exploratory data analysis (EDA) could be the only computationally feasible way of discovering complementary feature-sets under a MCC setting. How to efficiently explore becomes an essential computational issue. On top of this exploring issue, another issue is: how to synthesize pattern information resulted from complementary feature-sets. To resolve both issues and many others arising in later part of this paper, we devise a somehow central principle: {\bf Employing unsupervised machine learning to help us extract hidden categorical patterns and structures wherever we can on the process of exploring and computing information content}. The key is that categorical patterns and structures are to trim away redundant and defusing like complexity, and make computational explorations feasible and simultaneously make computed pattern information visible and explainable. This is the chief principle underlying our categorical exploratory data analysis (CEDA).
	
Our CEDA works without imposing unrealistic assumptions or man-made structures. On top of building a $K\times K$ mutual conditional entropy matrix, we can rank all the features with respect to their directed association to the categorical variable of pitcher-ID. The ranking queue and the small number of synergistic feature-groups collectively form a road-map for us to carry out exploratory computing task efficiently and precisely with large $K$.
	
Further, upon a mixing geometry of all point-clouds specified by a feature-set, CEDA goes on to compute one label embedding tree and one predictive map. Our label embedding tree is computed based on stochastic partial ordering of ``an unspecified distance'' among all point-clouds \cite{hsiehwang}. This indirect ``distance'' measure turns out to be realistic and robust. The resultant tree explicitly reveals a global geometry upon the label space, and facilitate serial binary competitions that are carried out by descending from the tree top to its bottom. Such a series of competitions is an essential new development here because each testing data point leads us to a possibly asymmetric mixing geometric pattern category.

The computed list of mixing geometric pattern categories collectively presents a visible label-specific predictive map or pie-chart. From MCC information content perspective, such a list of categories explicitly reveals which part of label-specific point-cloud can be perfectly predicted via a singleton-label, and which parts are predicted by which label-sets. Indeed each category prescribes an error-free decision from one perspective of one feature-set. Thus, when we arrange two feature-sets into an ordered chain, the second feature-set is used to dissect the uncertainty pertaining to the first feature-set. And so on, we can form a chain of two, three or more feature-sets. This is a way of making use of complementary feature-sets to exploring local mixing information. That is how the diversity in MCC information content is constructed. Also it is why we have a basis for perfect predictions with perspective, and at the same time can map out uncertainty resulted from popular machine learning algorithms.

Thus, our MCC information content must allow that all our decision-makings are fully supported by data. This is a fundamental standpoint for Data Science regarding inferences, which is only a small part of data analysis. In contrast, in statistics and machine learning literatures \cite{allwein01,hastie01,rifkin04}, inference is the primary focus and is usually placed on selecting the best feature-set with respect to a man-made criterion, such as minimizing the predictive error rate. Such inference, foremost, limits us to a singleton based prediction. That is, this criterion indeed forces us to make mistakes even when a singleton-prediction is not supported by data.
		
We illustrate our CEDA developments and MCC information content under two MCC settings: slider and curveball.

\subsection{Illustrating CEDA on Slider data}
We use a slider dataset consisting of only 5 MLB pitchers from the 2017 season to illustrate our CEDA computing. Here we take a MLB pitcher-ID as a label onto his own slider pitches. Our goal is to extract full MCC information content, which is defined as the collective of mixing characteristics revealed in geometries in panels-A\&B of Fig. \ref{fig:infogeoS} and many other potential feature-sets.

We reiterate that slider's speed is slightly less than fastball, but much higher than curveball within a pitcher's pitching repertoire. Its spin direction (``spin{\_}dir'') has a much wider range than that of both fastball and curveball. That is, due to Magnus effect, its horizontal movement (``pfx{\_}x'') is very versatile. A pitcher does vary spin direction and rate and speed to create a large range of spectrums of ``pfx{\_}x'' and vertical movement (``pfx{\_}z'') to deal with batters. So, a good slider pitcher is a wizard of spin direction and rate and speed.

The Slider's mutual conditional entropy matrix, as shown in panel-A of Fig. \ref{fig:MCEM}, clearly reveals 5 evident blocks as marked along its diagonal. These 5 highly associated feature groups with memberships listed as follows: Group-A: \{``end{\_}speed'', ``start{\_}speed'', ``vyo''\}; Group-B: \{``px'', ``x'', ``vz0'', ``pz''\}; Group-C: \{``z0'', ``vx0'', ``x0'' \}; Group-D: \{``spin{\_}dir'', ``break{\_}angle'', ``ax'', ``pfx{\_}x''\}; Group-E: \{``spin{\_}rate'', ``break{\_}length'', ``pfx{\_}z'',``az''\}. If we take each synergistic feature-group as a mechanical entity, we have achieve the ``dimension reduction'' from $K=21$ to 5. Via directed associations from pitcher-ID, we find the first 6 highest associative features (in increasing order) are: \{``x0'', ``z0'', ``spin{\_}dir'', ``vx0'', ``ax'', ``pfx{\_}x''\}.

Our primary explorations in CEDA take two routes: 1) explore the 5 the feature-groups and their combination; 2) or explore increasing series of feature-sets based on the ranked directed associations. After identifying several potential features, our explorations turn into tinkering by adding a-feature-at-a-time from other potential feature-sets.
	
The five computed synergistic feature-groups as five feature-sets give rise to five rather distinct mixing geometries of five pitchers' point-clouds. To precisely extract and transparently display any mixing geometry with respect to a given feature-set, our CEDA developments adopt the following stepwise protocol: the whole data set is divided into one major training data subset (80\%) and one minor testing data subset(20\%),
	\begin{description}
		\item[[MCC-Q1]]: First, based on the training data subset, we prescribe one genuine computational concept to derive relative closeness among all label-specific point-clouds given a feature-set, and then build a label embedding tree upon the label space;
		\item[[MCC-Q2]]: Second, upon each data-point in the testing data subset, we construct a serial binary $K$-nearest neighbor competitions between the right-and-left branches descending from the top of the binary structure of the label embedding tree, and we design a stopping-scheme for such a series of binary competitions. Upon each pitcher' testing data, we build a ``pie-chart'' to represent resultant local mixing geometric patterns. Collecting these pie-chart displays across all pitchers and construct a matrix representation for a list of categories of local mixing patterns;
		\item[[MCC-Q3]]: Third, by exploring along the two aforementioned routes, we choose an ordered chain of feature-sets: the latter feature-set is set to expose uncertainty of local mixing geometric patterns resulted from the former feature-set. Throughout the chain explorations, we present a collection of mixing geometric pattern categories of various orders in a series of ordered tables. Ideally, the proportion of testing data points falling into mixing geometric categories of ``certainty'' is as large as possible.
	\end{description}

\subsubsection{Computing a Label embedding tree (LET)}
For [MCC-Q1] in our CEDA protocol, it is essential that dimensionality of point-cloud becomes a key factor for any mathematical definition of direct distance measure. For instance, recently researchers apply Optimal Transport (OT) \cite{solomon18}, Gromov-Hausdorff or Gromov-Wasstein distances to evaluate a distance between two point-clouds \cite{memoli04,memoli11}. Since such a direct measure between two point-clouds often is based on a pair of distribution functions. Not only the curse of dimensionality will apply here, but also their non-robustness to shapes and tails behaviors of distributions will jointly worsen their effectiveness. Thus, their uses in evaluating distances among diverse mixing geometries are rather limited from practical and realistic point of views. To avoid such shortcomings, we turn to the concept of relative-distance or relative-closeness.
	
Here a relative-distance among point-clouds is a deduced one, not directly measured ones.  All directly measured distances are unrealistic anyway due to varying shapes. The most critically needed global structure, such as a tree, on label space doesn't strictly required a directly measurable distance, see detailed developments and experiments in \cite{hsiehwang} and an algorithm given below.

\begin{figure*}[t]
		\centering	
		\includegraphics[width=2.5in]{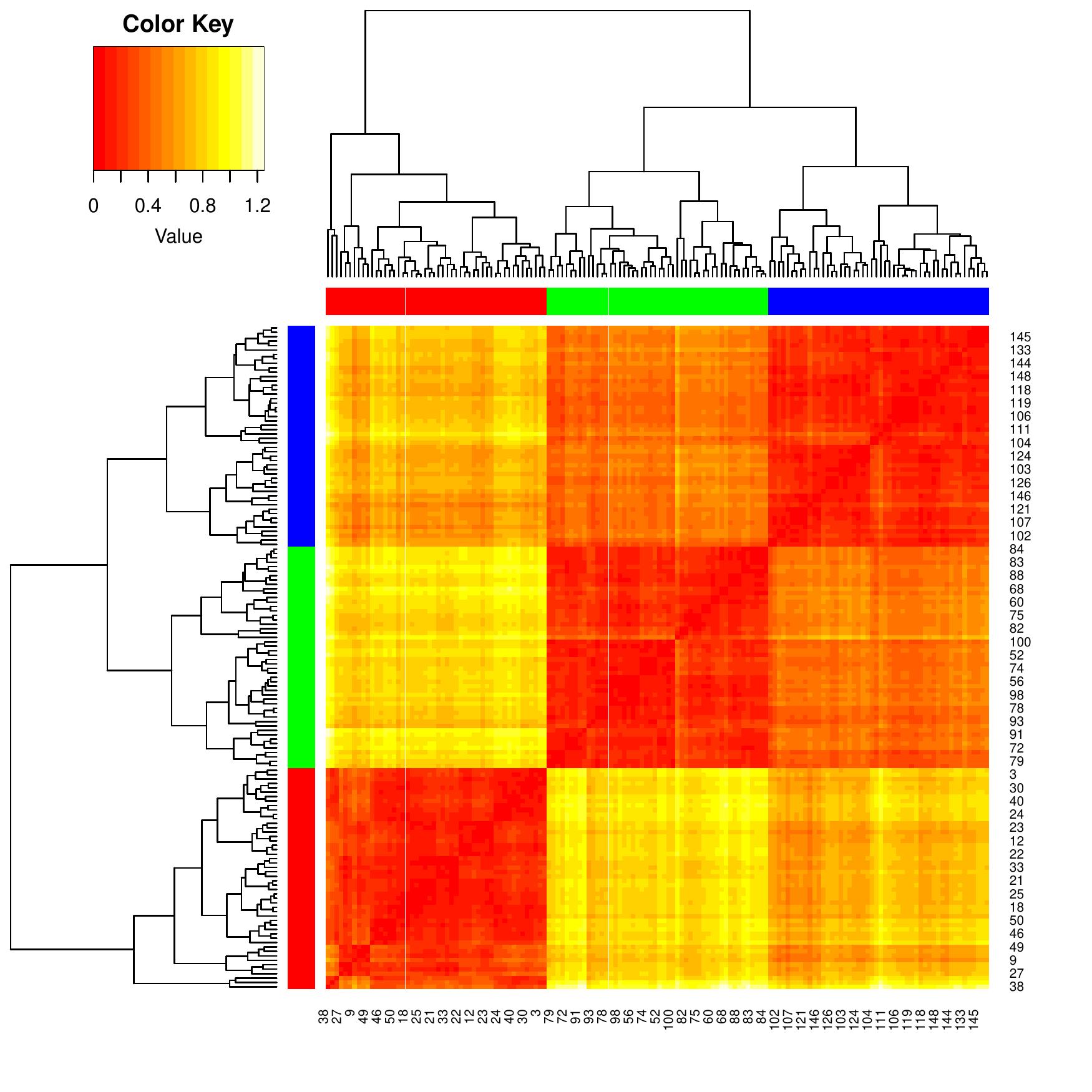}
	\includegraphics[width=0.5\textwidth]{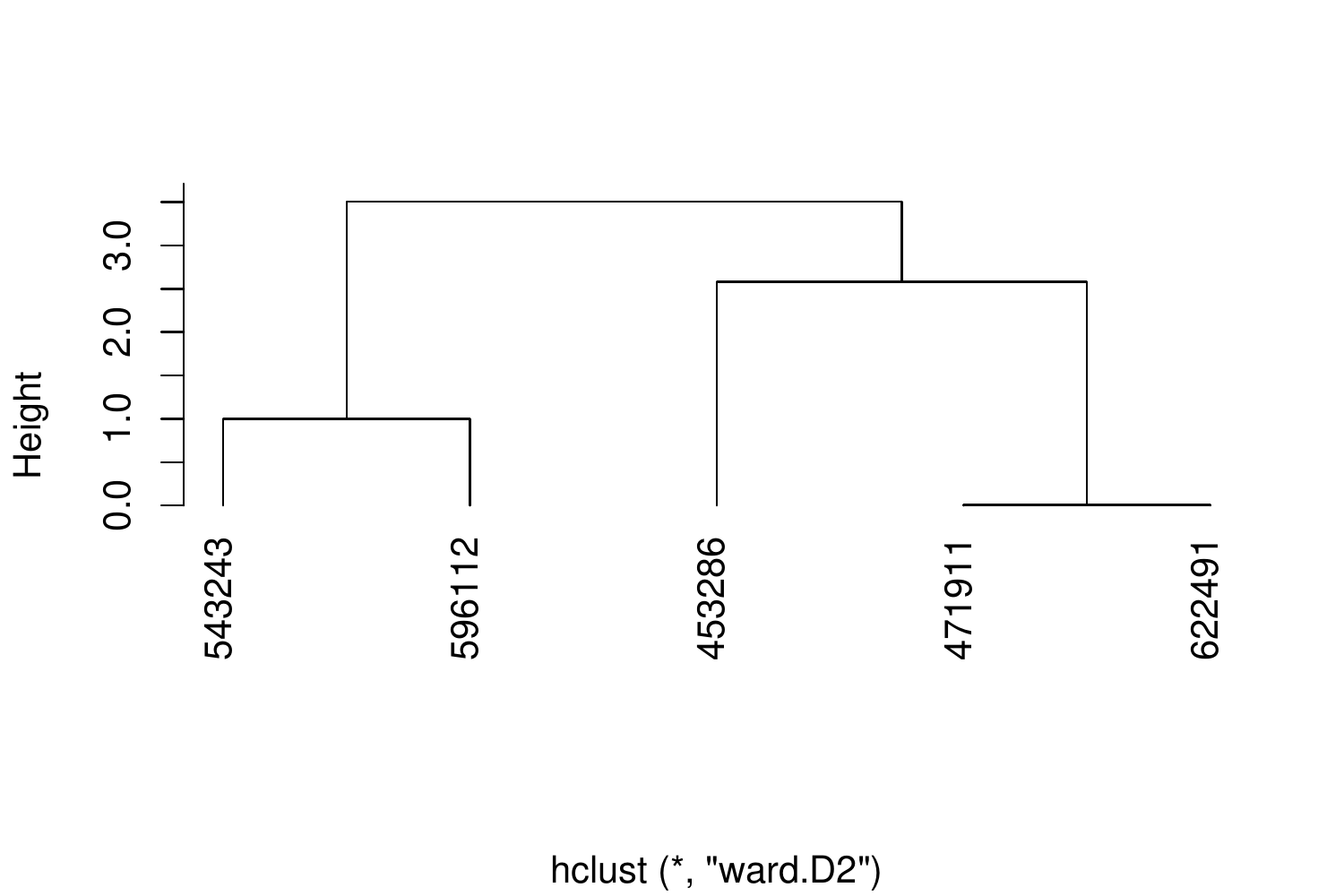}
		\caption{Global geometry of point-cloud based on Feature-group-C :(A) Distance matrix of three sets of randomly selected data points from three pitchers: 453286(red), 543243(green) and 596112(blue); (B) Label embedding tree.}
		\label{fig:figure2}
\end{figure*}

Here we only briefly go over the underlying ideas and steps. To creating indirect data-driven measurements of relative-distance among point-clouds, we perform a unsupervised machine learning on sampled slider pitches from three pitchers (as color-coded) with respect to a feature-set of choice, as shown in the right panel of Fig. \ref{fig:figure2} for feature-group C. The resultant hierarchical clustering (HC) tree on the row axis of the heatmap, an Euclidean distance matrix of sampled data points, reveals that pitchers in green and blue are closer to each other than to the pitcher in red with respect to the sampled data points.

We interpret that this HC tree collectively aggregates many pieces of information of partial orderings: $dis(green, blue) < min(dis(green, red), dis(blue, red))$ without the need of knowing the ordering between $dis(green, red)$ and $dis(blue, red)$ with $dis(,)$ denoting the unspecified and undefined ``relative-distance''. Thus, if we repeatedly sample from the triplet of pitchers, and collect their partial ordering information, then collectively can discover a precise tree structure on the triplet of pitchers. By replicating same construction on all possible triplets of pitchers, we can arrive at a so-called label embedding tree (LET) on the label space, see panel-B of Fig. \ref{fig:figure2}. We propose the following algorithm to resolve [MCC-Q1].
	
	\paragraph{[Algorithm for label embedding tree (LET)]}
	\begin{description}
		\item[[T-1]] Given a set of $k$ features, choose one triple of labels (pitchers) at-a-time, and sample one triple of $k$-dim vectors from the three distinct labels' training point-clouds and evaluate three pairwise Euclidean distances. We only record binary partial ordering: which label-pair's pairwise-distance is dominated by the other two label-pairs' pairwise-distances;
		\item[[T-2]] Repeat Step-[T-1] many times across many triples of labels, and summarize and arrange all binary partial orderings of pairwise-distance-dominance into a dominance matrix with all possible label-pairs being arranged along its row and column axes;
		\item[[T-3]] Upon such a dominance matrix, which designates a column-pair dominating a row-pair, then each row-sum indicates the relative-closeness of the pair of labels among all possible pairs, while each column-sum indicates the ``relative-distance'' of the pair of labels;
		\item[[T-4]] The collection of column-sums forms a ``relative-distance'' matrix of all labels. A label embedding tree (LET) is built based on HC algorithm on this dominance-based relative-distance matrix.		
	\end{description}

\subsubsection{Developments of predictive map of mixing geometric pattern categories.}
The tree geometry of LET is one part of large-scale geometric information content. Due to its stochastic nature within a partial ordering based dominance matrix, the LET is very robust with respect to outlier and overlapping edges of point-clouds. Based on the binary structure of LET, we construct a tree-descending series of ``left-vs-right'' branch-to-branch competitions for each every testing data point. The serial results of competitions precisely indicate where to locate such a testing data point as it it is included in the entire mixing geometry of all point-clouds (based on training data). This is an efficient way of exposing all mixing geometric patterns involving in all $L$ point-clouds with respect to a chosen set of $k$ features. Such pattern information basically is invisible, or even hidden when $k >3$.
	
In regarding to [MCC-Q2] of our CEDA protocol, we propose the following algorithm to expose and extract local mixing geometric information, and then summary all results into a table called predictive map, which summarizes results from all label specific pie-chats. This table-format of map reveals middle-scale MCC information content in details.
	
\begin{figure*}[t]
		\centering
		\includegraphics[width=2.2in]{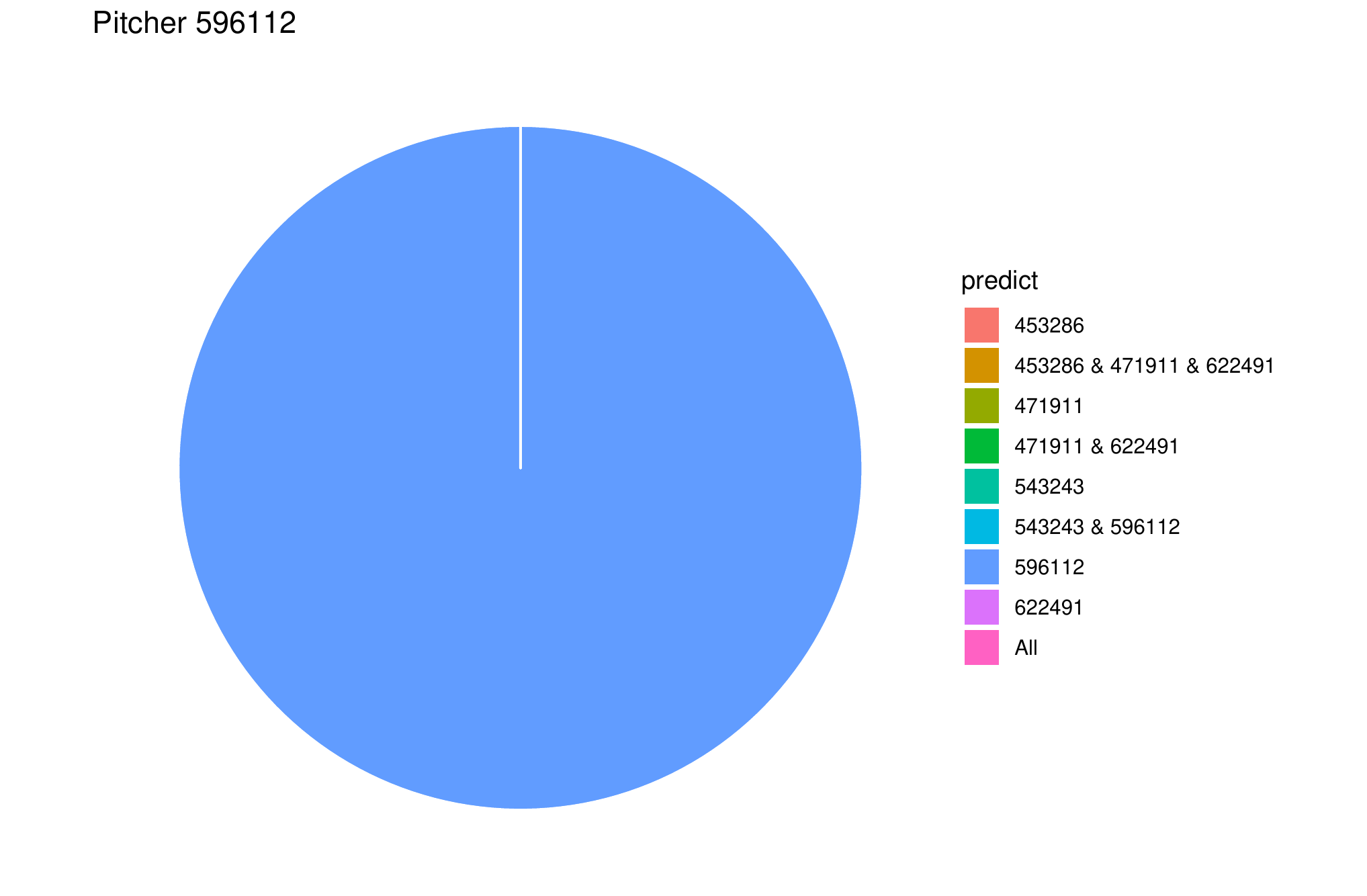}
		\includegraphics[width=2.2in]{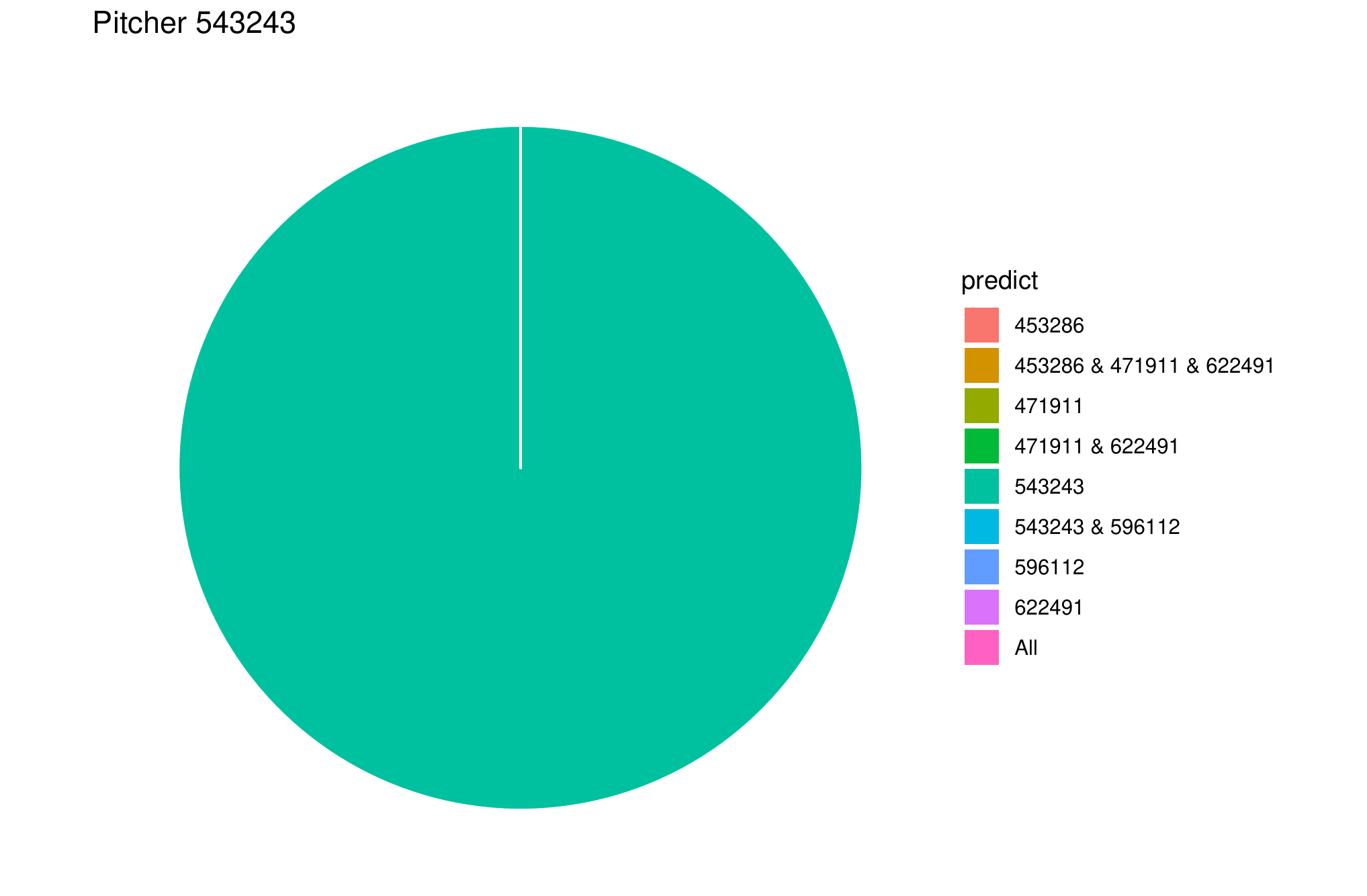}
		\includegraphics[width=2.2in]{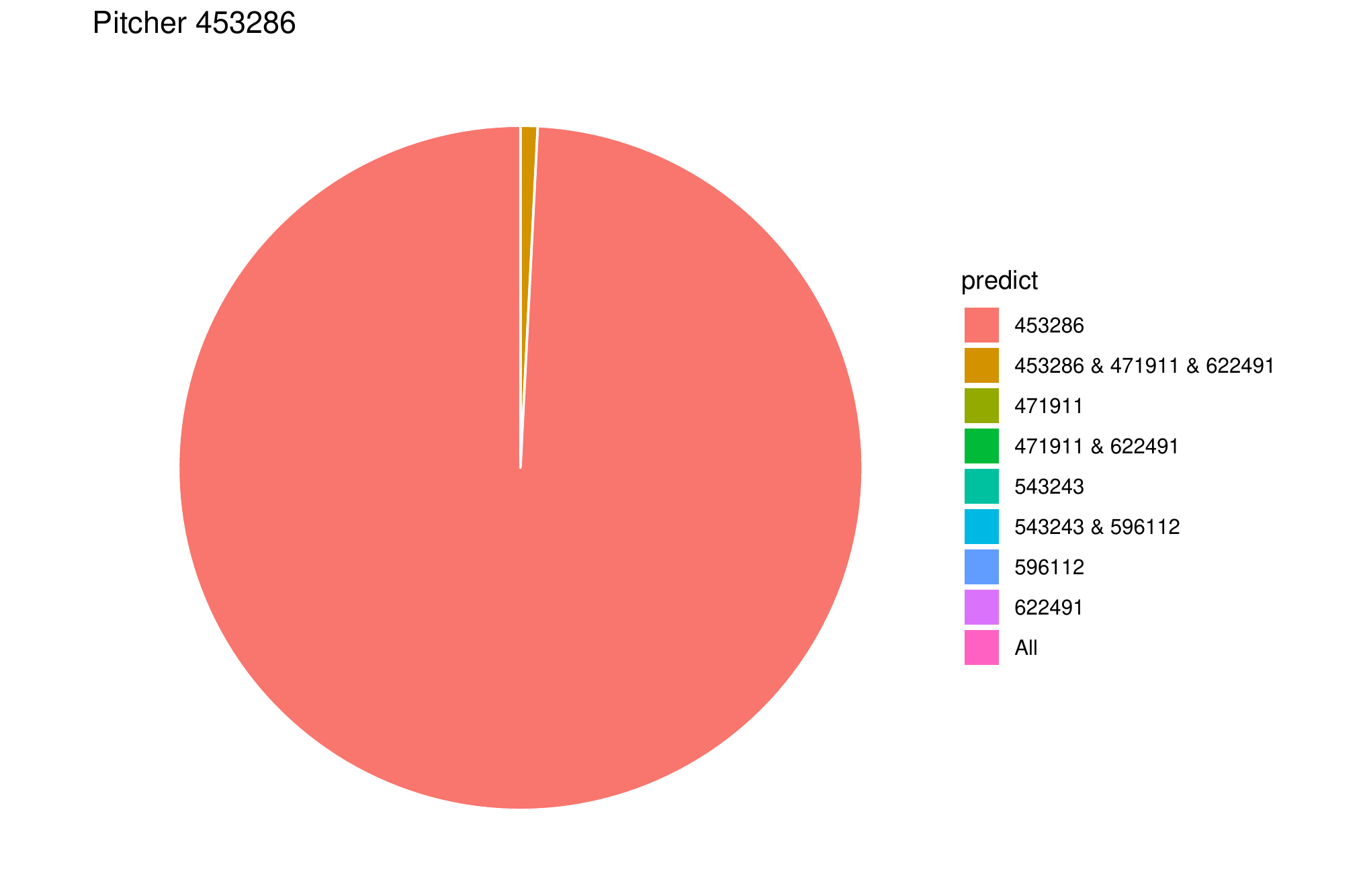}\\
		\includegraphics[width=2.2in]{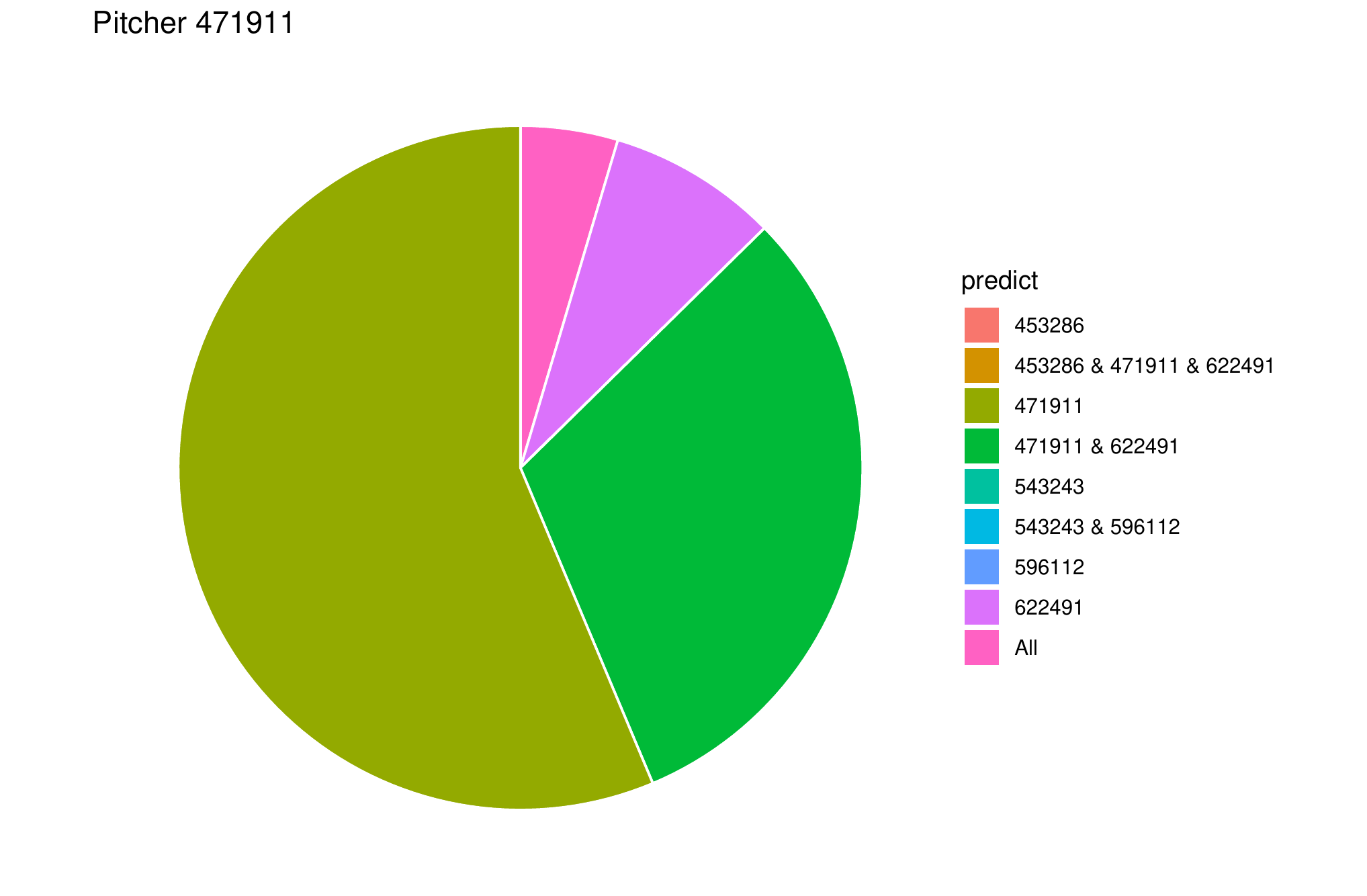}
		\includegraphics[width=2.2in]{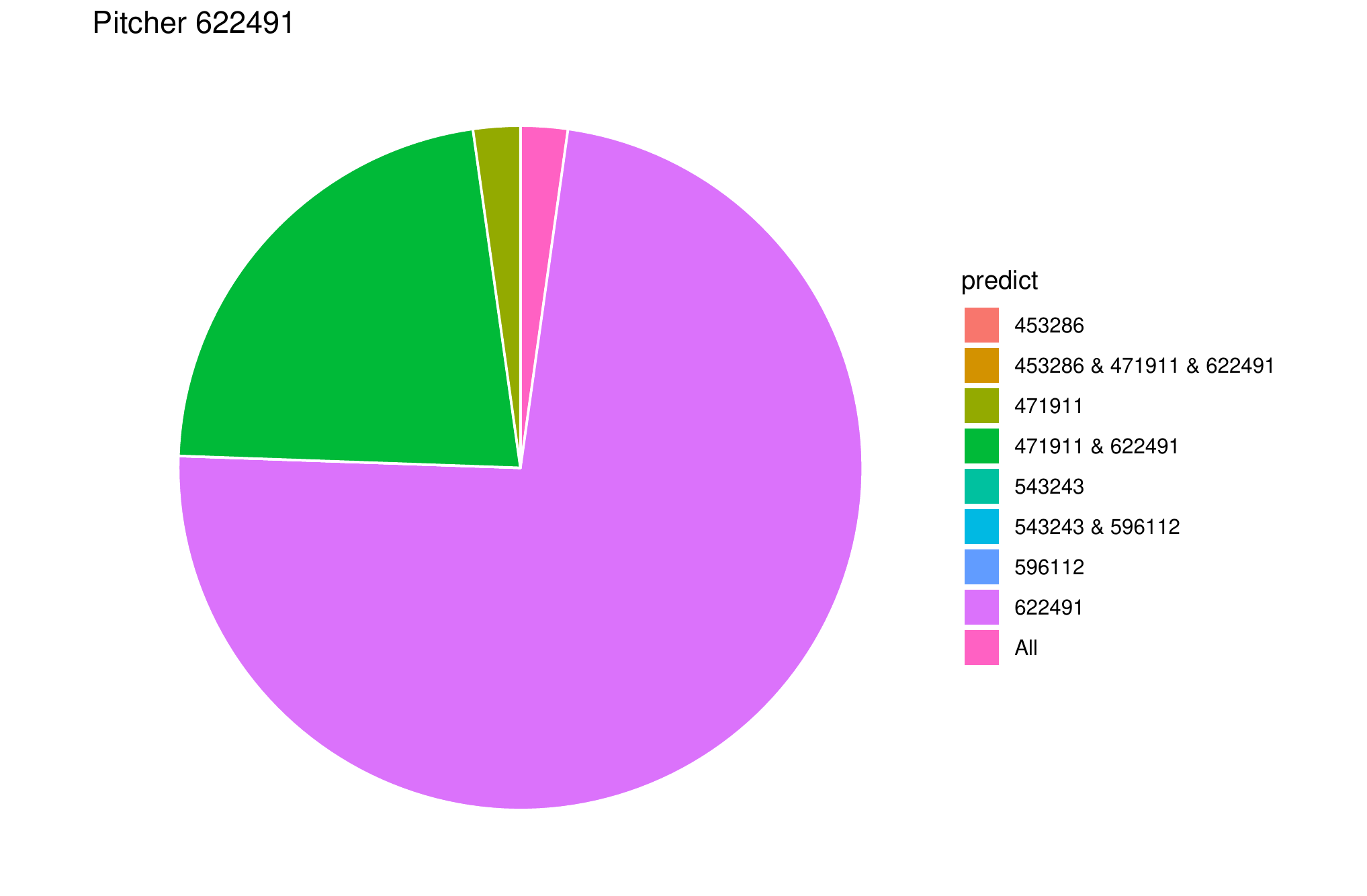}
		\caption{Predictive map of five slider pitchers based on feature-set-C with threshold range[0.65, 100/65] for pseudo-likelihood values via K(=20) nearest neighbors.}
		\label{fig:piechart}
	\end{figure*}

\begin{figure*}[t]
		\centering
		\includegraphics[width=0.33\textwidth]{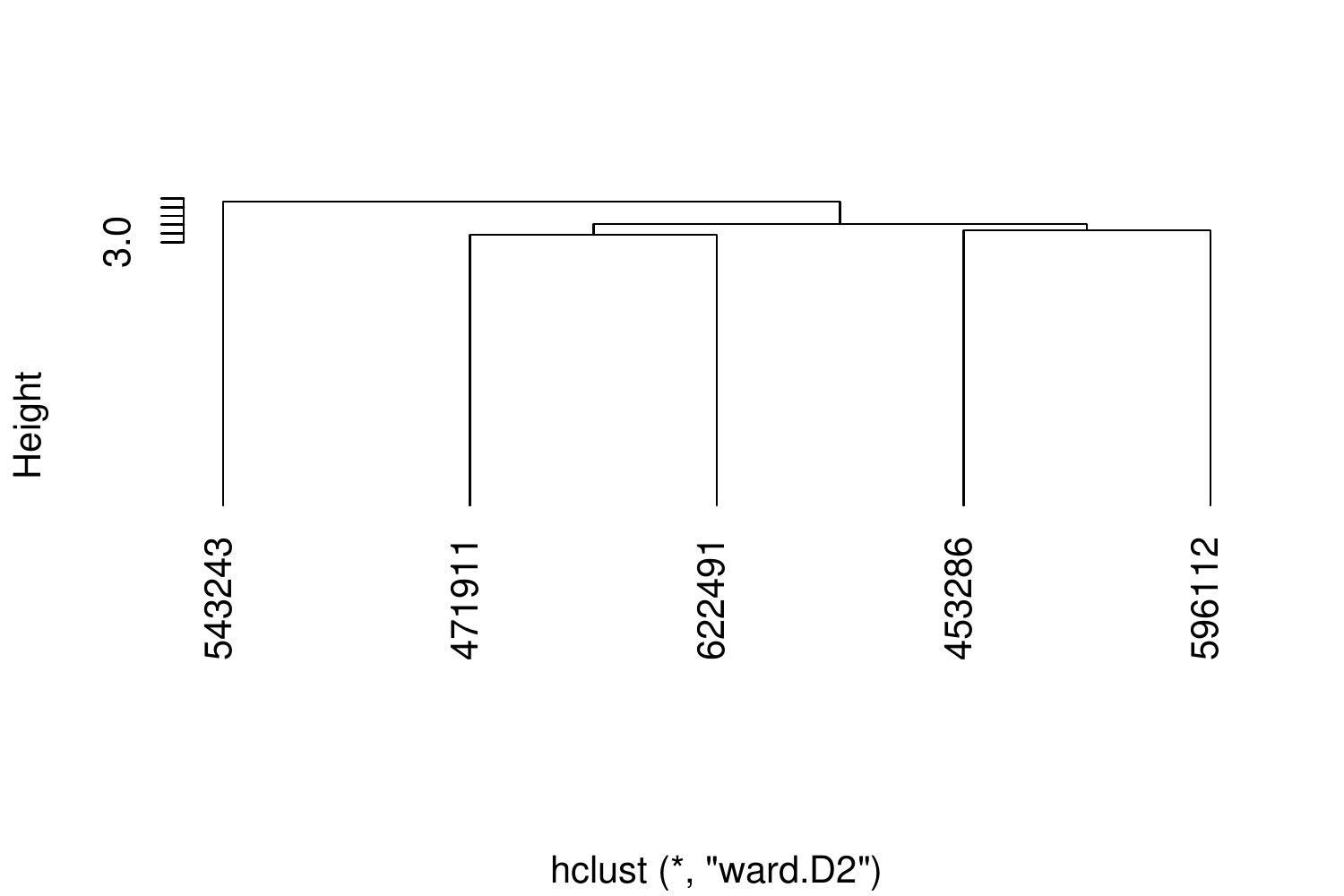}
		\includegraphics[width=0.33\textwidth]{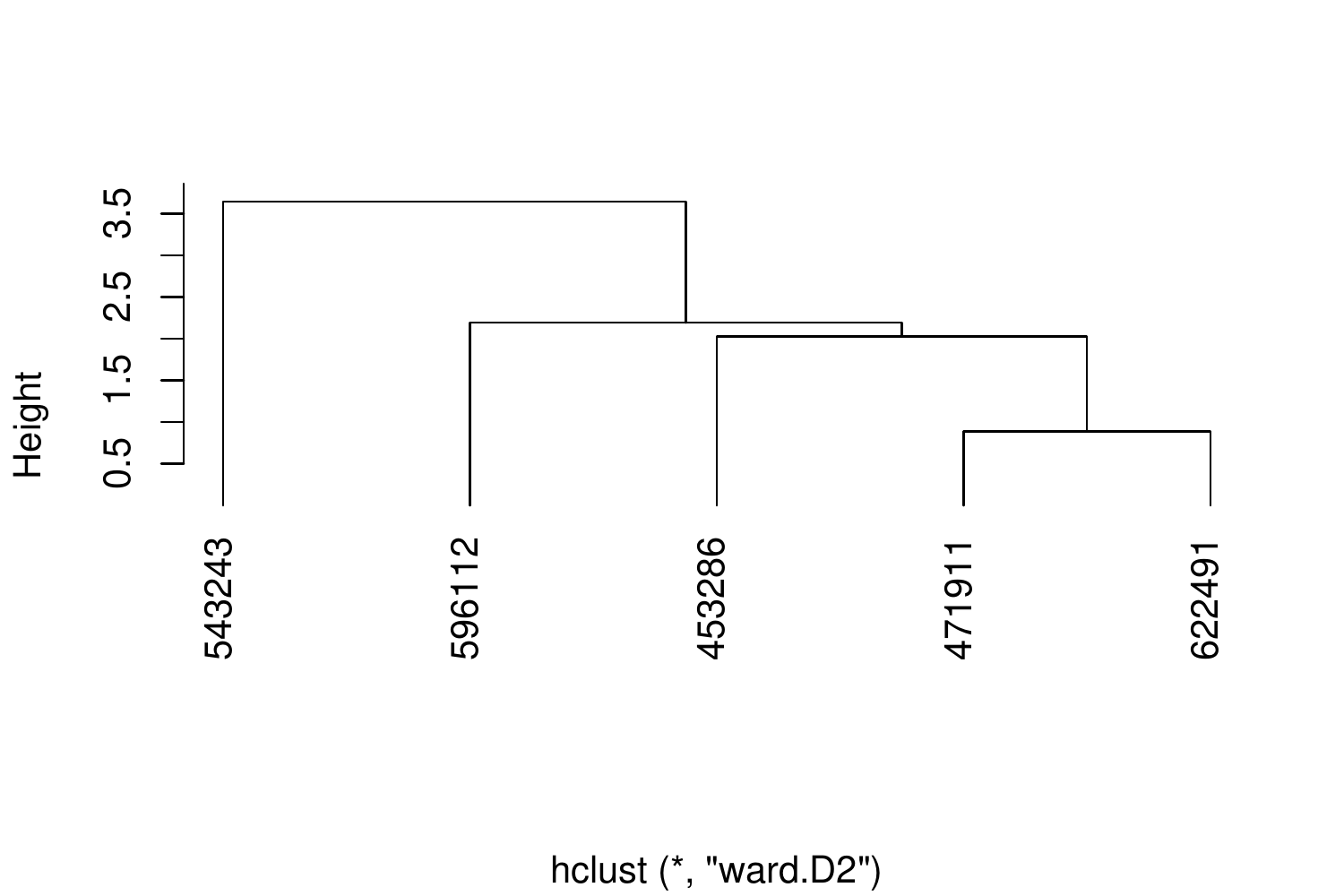}\\
		\includegraphics[width=0.33\textwidth]{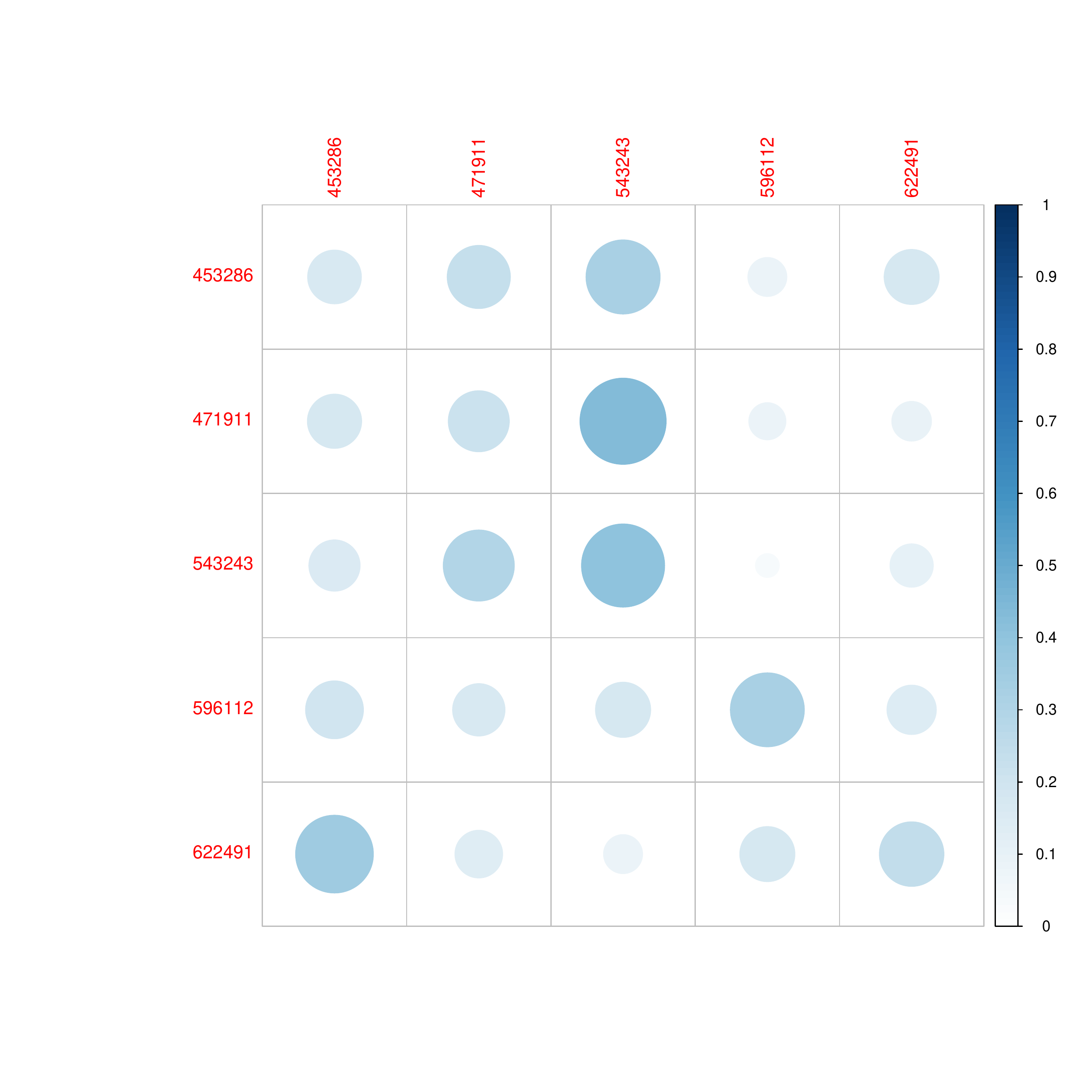}
		\includegraphics[width=0.33\textwidth]{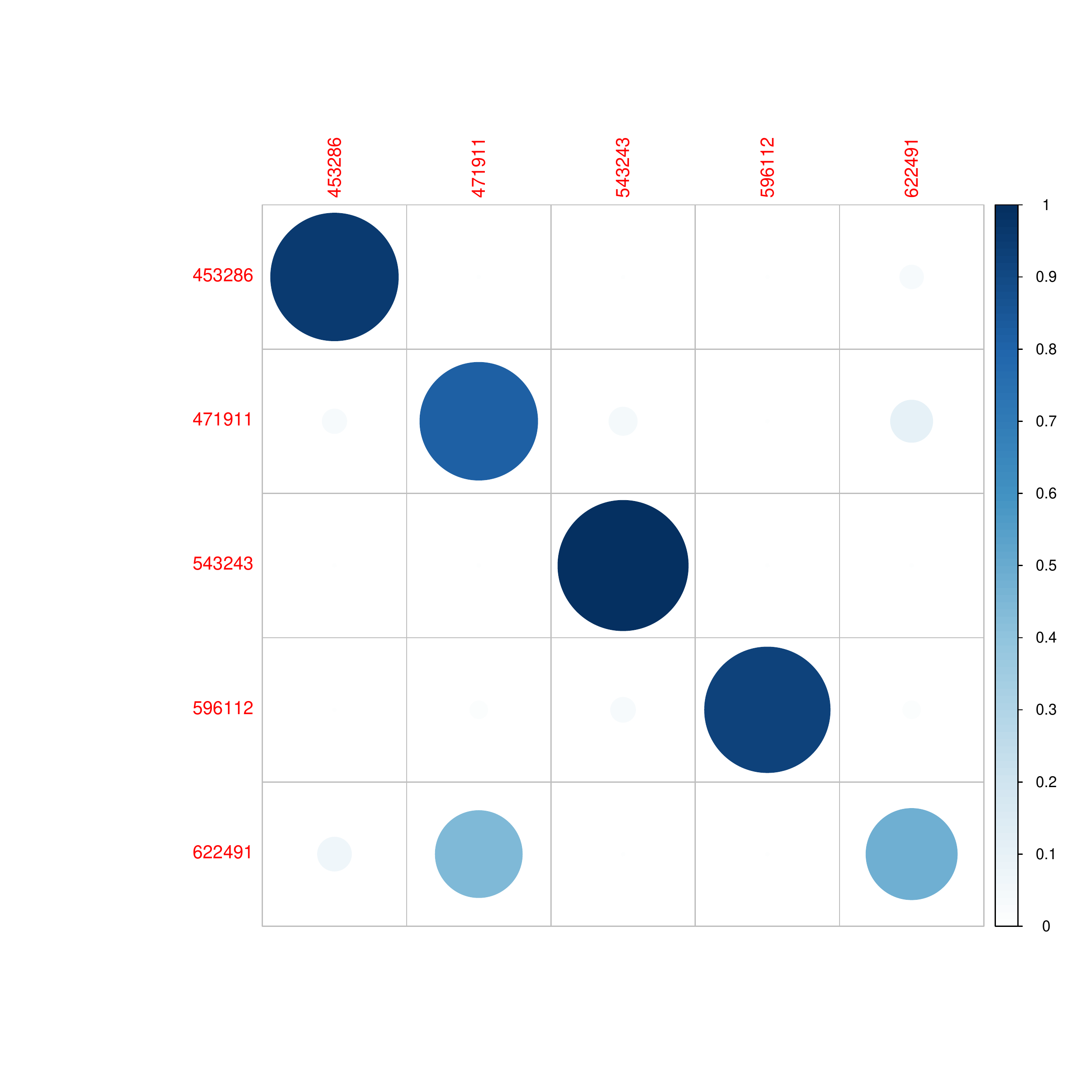}
		\caption{Slider's 2 label embedding tree and 2 predictive maps (with threshold value 1) with respect to feature-set-A(left column); and all 21 features(right column).}
		\label{fig:singleton}
	\end{figure*}

	\paragraph{[Algorithm for Predictive Map (PM)]}
	\begin{description}
		\item[[P-1]] Take a testing point-vector, say $x$, from the testing data set of a label. Starting from the internal-node on the very top of LET, perform binary (Left-branch vs Right branch) competition via $k^*-$nearest neighbors of $x$. Here $k^*=20$ is used. To decide which branch $x$ belongs, we separate the $k^*$ neighbors with respect to their Left-and right-branch memberships. We declare the winner according to the following policy: a) one branch has absolutely dominant in membership count; b) if both counts are not significantly different, then we build two membership-specific distance distributions. With each distribution, we calculate a Pseudo-likelihood (PL) value with respect to the median distance from $x$ to all members of the same branch. Then compare the ratio of the two PL-values with respect to a threshold range $[C_L, C_U]$: if the PL-ratio falls out of this range with clear winner being left- or right-branch, then we go on to the next step; if it falls within the range we stop the competition process and declare and record no-winner at the internal node;
		\item[[P-2]] Repeat the binary competition of Step-[P-1] at the winner branch's internal node and descend further down along the winning branches until the serial of binary games stop at one no-winner internal node, which can be a branch consisting of one singleton label. Then we record branch members of no-winner internal node as the predicted label-set, which could be a singleton, multiple or none. (The case of empty predicted label-set is a device of zero-shot learning for discovering outlier.)
		\item[[P-3]] Repeat Step-[P-1] and Step-[P-2] for all testing data points across all labels, and then partition each label's testing set with respect to all its members' predicted label-sets as categories in a form of a pie chart, as shown in Fig. \ref{fig:piechart}. By arranging all observed predicted label-sets across all labels along the row axis, and involving labels along the column axis, all predictions of testing data points from the testing data set are summarized into a matrix, which is called predictive map (PM), as would be seen in the Fig. \ref{fig:singleton} and the subsection below.
	\end{description}
[Here: Illustration example 1]

The five pie charts, shown in Fig. \ref{fig:piechart}, reveal very coherent and precise mixing geometric information among the five point-clouds. The two pie charts: pitcher-596112(purple) and pitch-543243(green), indicate that their point-clouds are well separated from all the rest of 3 point-clouds, respectively. Nearly so for pitch-453286(brown), except a small area of mixture involving three pitchers: \{453286 \& 471911 \&622491\}. It is very significant to note that this mixture of \{453286 \& 471911 \&622491\} is uniquely pertaining to pitcher-453286. No such mixture is found upon any other pitcher. This is a key piece of information content based on local scale mixing geometry. Thus, any testing data-point resulted in this pitcher-specific mixture must uniquely belongs to pitcher-453286, not the other two pitchers:471911 \&622491. {\bf We conclude that, from prediction point of view, pitches of pitcher-453286 can be perfectly predicted with 100\% precision in singleton format.} More examples of such asymmetry are reported below. This mixing geometric asymmetry based result is at odds with all existing works in statistics and machine learning literature.

Mixing geometric information of pitcher-471911 and pitcher-622491 is also diverse and characteristic. More than 50\% of pitches of pitcher-471911 are standing alone and identified as pitcher-471911, while there is about 75\% of pitches of pitcher-622491 are standing alone and identified as pitcher-622491. There are about 20\% pitches of pitcher-471911 and 15\% of pitches of pitcher-622491 being identified for both pitchers: \{471911 \& 622491\}, respectively. It is legitimate to claim that a testing data point falling into this mixture of \{471911 \& 622491\} is predicted as (471911, 622491) without choosing between the two because such a decision is fully supported by the mixing geometry. It is unnecessary, or even unnatural to choose one against the other. We need to include extra information computed from different feature-sets in order to separate between these two pitchers.

In both pie charts, there are visible proportions of pitches from both pitchers:471911 \& 622491, are identified ``wrongly''. The reason is that a pitch of pitcher-472911 is intensively surrounded by pitches of pitcher-622491, or vise versa. Since this mixing pattern is feature-set specific. Such a mixing pattern might be altered significantly with respect to different feature-sets. This is why we need to seek for complementary feature-sets.

In sharp contrast, if the choice of threshold range $[C_L, C_U]=[1,1]$, which is equivalent to using a threshold value 1 on PL-ratio, then we force ourselves to choose a winning branch against a losing branch at each internal node from the top of LET toward its very bottom. Consequentially, all predicted label-sets under such a scheme of thresholding are in a form singleton. It is essential to note that potential errors are to be accumulated, and at the same time the capability of declaring an outlier is discarded. In other words, we force ourself to make mistakes by disregarding how much information is indeed supported by the training data.

We illustrate results obtained under this thresholding scheme with two less informative feature-sets. Their LETs and predictive maps are respectively reported in the two column of panels of Fig. \ref{fig:singleton} for the feature-set-A and the feature-set of all features. From two collectives of resultant predictive maps via matrix-form, we evidently also see the asymmetry of mixing geometries of involving point-clouds. We also see that the feature-group of all features is rather more informative than the feature-set-A.

\subsection{Slider's MCC information content in full}
So far we can compute a label embedding tree (LET) and a predictive maps (in pie chart format) for any feature-set. The next task is: how to find and see the MCC information content in full. With the discovery of mixing asymmetry being prevalent in mixing geometries, we recognize the great potentials of complementary feature-sets. That is, we aim at constructing a chain of feature-sets to extract all possibly viable patterns from an ordered series of mixing geometries. From this perspective of serial mixing geometries, which is in certain sense in accord with the boosting concept in machine learning, our task of computing MCC information content in full is rather unique in the sense of being distinct from the classic issue of selecting the most informative feature-set.

For expositional conciseness we also encode pitcher-IDs as follows: ``a(453286)'', ``b (471911)'' ``c(543243)'', ``d(956112)'' and ``e(622491)''. Our explorations lead us to the feature group C when the choice of threshold range $[C_L, C_U]=[0.65, 100/65]$ for PL-ratio is adopted. The five resultant pie-charts for the five pitcher-labels, as reported in Fig. \ref{fig:piechart}, bring out the interesting fact: a distinct and unique pitcher-specific local mixing category, not sharing with any other pitchers, certainly identifies the pitcher.  On top of being distinct to classic predictive classification, this recognition becomes a guideline for collecting pieces of MCC information content.
		
For the last step, [MCC-Q3], of protocol of CEDA, we explore fine scale MCC information content by exploring a potential chain of feature-sets. We choose to begin with the feature-group C. Its five pie charts are converted into a table, as shown in the left-panel of Table~\ref{tab:tab1}, with respect to 8 ``observed'' mixing geometric pattern categories arranged and listed on its row axis. In this fashion, MCC information content from feature-set-C is clearly revealed and organized. The three pitchers \{a, c, d\} are perfectly separated with respect to their own pitcher-specific pattern-categories. The rest of 4 mixing geometric pattern-categories contains pitches from both pitchers-b and -e, which are marked by ``*''. Such a pattern-category sharing means a locality of``uncertainty'' among involving pitchers. For instance, the category \{*b\} has 49 pitches from pitcher-b and 1 pitch from pitcher-e, while the category \{*e\} has 7 pitches from pitcher-b and 33 pitches from pitcher-e.

We decide to further dissect such uncertainty by looking into these localities through a different perspective via a feature-set: \{``x0'', ``z0'', ``spin{\_}dir''\}, which has its pie chart being summarized and presented in right panel of Table~\ref {tab:tab1} with 7 categories more uncertainty. In Table~\ref {tab:tab2}, the 4 ``*'' marked mixing geometric pattern categories via feature-set-C are projected with respect to 6 mixing geometric pattern-categories (without \{a\}). From its first three columns for \{*b\}, we see that the 49 pitches of pitcher-b are exclusively divided into 3 second-order categories: 34 in \{*b-b\}, 14 in \{*b-be\} and 1 in \{all(abcde)\}, and 1 pitch from pitcher-e is alone in \{*b-e\}. Therefore, these 50 pitches are exclusively located. This is a sense of complementary feature-sets.

In contrast, the 40 pitches in category \{*e\} are divided into 2 second-order categories: 25 in \{*e-e\} (with 1 pitch belonging to pitcher-b and 24 belonging to pitcher-e) and 15 in \{*e-be\} (6 from pitcher-b and 9-from pitcher-e). Therefore, these two second-order categories are subject to uncertainty and need further explorations from another perspective of a new feature-set. Likewise, second-order categories derived from 1st-order categories \{*be\} and \{*all\}. We choose the feature-set-A\&B\&C, including members of three feature groups: A, B and C, as the 3rd in our chain of feature-sets. Pattern-categories of mixing geometry of feature-set-A\&B\&C are given in Table~\ref {tab:tab3}.

All the 3rd-order mixing geometric pattern-categories are listed in Table~\ref {tab:tab4}. This table shows very promising results. There are 15 out of 21 triplet categories obtaining certainty in regarding to being exclusively in pitcher-b or pitcher-e. This phenomenon manifests clear characters of 3 complementary feature-sets in a chain.

As for pitches belonging to the rest of 5 categories of 3rd-order, geometrically speaking, intensive mixing between the two pitcher-b and pitcher-e are found within these localities. Therefore, it is reasonable as well as necessary to make prediction decision as \{pitcher-b, pitcher-e\} within these localities. On the other hand, making any singleton-based decision upon such localities is strongly against the evidence supported by the data. Surely, odds-ratios should company with the predictive label-sets.

In summary, with a carefully selected chain of 3 feature-sets, the MCC information content of slider data set is explored in depth and collectively represented by its LET and a series of ordered predictive maps. There are at least 3 notable merits of such MCC information content. The chief one is that the 1st, 2nd and 3rd order of mixing geometric pattern categories jointly provide the basis for understanding the rationales underlying the labeling. The second merit is that they offer a platform for error-free decision-making, that is, all predictive results are supported by the data. We demonstrate the same characters of information content on curveball data set in the next subsection. The third merit is that such a serial of tables of mixing geometric pattern categories allow us to dissect the uncertainty of unexplainable black-boxed results derived from popular machine learning algorithms, such as random forest and various boosting approaches. This merit is somehow surprising as would be demonstrated after the next subsection,.

\begin{table*}[!htb]
		\caption{Predictive maps of two feature-sets: C (left); \{``x0'',``spin{\_}dir'',``z0''\} (right).}
		
			\begin{tabular}{|l|lllll||l|lllll|}\hline
				feature-set-C& a   & b  & c  & d  & e &\{x0, spin{\_}dir z0\}& a   & b  & c  & d  & e  \\ \hline
				a              & 123 & 0  & 0  & 0  & 0  &	a             & 124 & 0  & 0  & 0  & 0  \\
				*b             & 0   & 49 & 0  & 0  & 1  & 	b             & 0   & 43 & 0  & 0  & 2  \\
				c              & 0   & 0  & 91 & 0  & 0  & c             & 0   & 0  & 85 & 1  & 0 \\
				d              & 0   & 0  & 0  & 56 & 0  &	 d             & 0   & 0  & 0  & 52 & 0\\
				*e             & 0   & 7  & 0  & 0  & 33 &	e             & 0   & 3  & 0  & 0  & 26 \\
				a, b, e        & 1   & 0  & 0  & 0  & 0  &	b, e          & 0   & 40 & 0  & 0  & 17  \\
				*b, e          & 0   & 27 & 0  & 0  & 10 &	a, b, c, d, e & 0   & 1  & 6  & 3  & 0  \\
				*all        & 0   & 4  & 0  & 0  & 1 &                 &     &     &     &     &     \\\hline
				
			\end{tabular}
            \label{tab:tab1}
\end{table*}

\begin{table*}[!htb]
\caption{Predictive results of chain: feature-sect-C to feature-set \{``x0'',``spin{\_}dir'',``z0''\}.}

				\begin{tabular}{|l|ll||l|ll||l|ll||l|ll|} \hline
					*b	&b	&e	&	*e	&b&	e&		*be&	b	&e&		*all&	b&	e\\ \hline
					b&	34&	0&  b	&	0&	0&  b	&   6&	2&	b	&	3&	0\\
					c&	0&	0&	c	&	0&	0&	c	&	0&	0&	c	&	0&	0\\
					d&	0&	0&	d	&	0&	0&	d	&	0&	0&	d	&	0&	0\\
					e&	0&	1&  e	&	1&	24& e	&	2&	1&  e	&	0&	0\\
					be&	14&	0&  be	&	6&	9&	be	&	19&	7& be	&   1&	1\\
				   all& 1&	0&	all	&	0&	0&	all	&	0&	0&	all	&	0&	0\\\hline
			\end{tabular}
            \label{tab:tab2}

		\end{table*}

\begin{table*}[!htb]

\caption{Predictive map of feature-set: A\&B\&C.}

			\begin{tabular}{|l|lllll|}\hline
				 ABC-&a   & b  & c  & d  & e \\ \hline
						a             & 122 & 0  & 0  & 0  & 0  \\
				 		b             & 0   & 34 & 0  & 0  & 1  \\
				 	    c             & 0   & 0  & 89 & 0  & 0  \\
					    d             & 0   & 2  & 1  & 56 & 0  \\
						e             & 0   & 0  & 0  & 0  & 15 \\
						abe           & 2   & 0  & 0  & 0  & 4  \\
						be            & 0   & 44 & 0  & 0  & 19 \\
				    	all(abcde)    & 0   & 7  & 1  & 0  & 6 \\\hline
				
			\end{tabular}
\label{tab:tab3}

\end{table*}

\begin{table*}[!htb]
\caption{Predictive results of chain: feature-sect-C to feature-set \{``x0'',``spin{\_}dir'',``z0''\} to feature-set A\&B\&C.}
\label{tab:tab4}

				\begin{tabular}{|l|ll|}
					\hline
					Chain of 3 sets	&b	&e\\\hline
					e-e-b&	1&	1 \\
					e-be-b&	1&	0 \\
					e-e-abe&	0&	4\\
					e-e-be&	0&	6\\
					e-be-be	&5&	5\\
					e-e-all&	0&	1\\
					e-be-all&	0&	1\\
					be-b-b&	4&	0\\
					be-e-b&	1&	0\\
					be-be-b&	4&	0 \\\hline
				\end{tabular}						
				\begin{tabular}{|l|ll|}
					\hline
					Chain of 3 sets (cont'd)	&b	&e\\\hline
					be-be-d&	1&	0\\
					be-b-be&	2&	1\\
					be-be-be&	12&	6\\
					be-b-all&	0&	1\\
					be-e-all&	1&	1\\
					be-be-all&	2&	1\\
					all-b-b&	1&	0\\
					all-be-be&	1&	0\\
					all-b-all&	2&	0\\
					all-be-all&	0&	1\\\hline
				\end{tabular}

		\end{table*}

\subsection{Curveball's MCC information content}
We report here our CEDA computational results under the MCC setting of curveball data set. We demonstrate that the same CEDA computing and exploring theme works for curveball's MCC information content. The selected chain of feature-sets turns out to be rather efficient in this MCC setting.

Curveball is a pitch-type with top-spin, which is the opposite of fastball's back-spin. Unlike fastball, generating top-spin is a bit unnatural. So, a curveball pitch in general is much slower than fastball and  slider for any MLB pitcher. Its characteristic trajectory has a significant vertical drop (measured by ``pfx{\_}z'') when reaching the home plate. This drastic drop, which is caused by Magnus effect on top of the gravity force, makes this pitch-type rather effective in dealing with batters. However, in sharp contrast with slider, its horizontal movement (measured by ``pfx{\_}x'') has a narrow range at the relatively small end.
		
Curveball data set consists of 7 $(=K)$ MLB pitchers. We compute its $21 \times 21$ mutual conditional entropy matrix, as shown in panel-B of Fig. \ref {fig:MCEM}. This heatmap reveals 6 evident blocks as marked along its diagonal. These 6 highly associated feature groups are: Group-A: \{``px'', ``x''\}; Group-B: \{``vz0'', ``pz''\}; Group-C: \{``spin{\_}dir'', ``break{\_}angle'', ``ax'', ``pfx{\_}x''\}; Group-D: \{``break{\_}length'', ``pfx{\_}z'',``az''\}. Group-E: \{``end{\_}speed'', ``start{\_}speed'', ``vyo''\}; Group-F: \{``ay'', ``spin{\_}rate'', ``z0'', ``vx0'', ``x0'' \}. We also calculate the directed associations of the 21 features toward the label space and rank them. The first 6 highest associations (in increasing order) are: \{``x0'', ``z0'', ``spin{\_}dir'', ``start{\_}speed'', ``vy0'', ``break{\_}angle''\}.  This feature-set of 6 is denoted as BB.
		
\begin{figure*}[h!]
			\centering
			\includegraphics[width=2in]{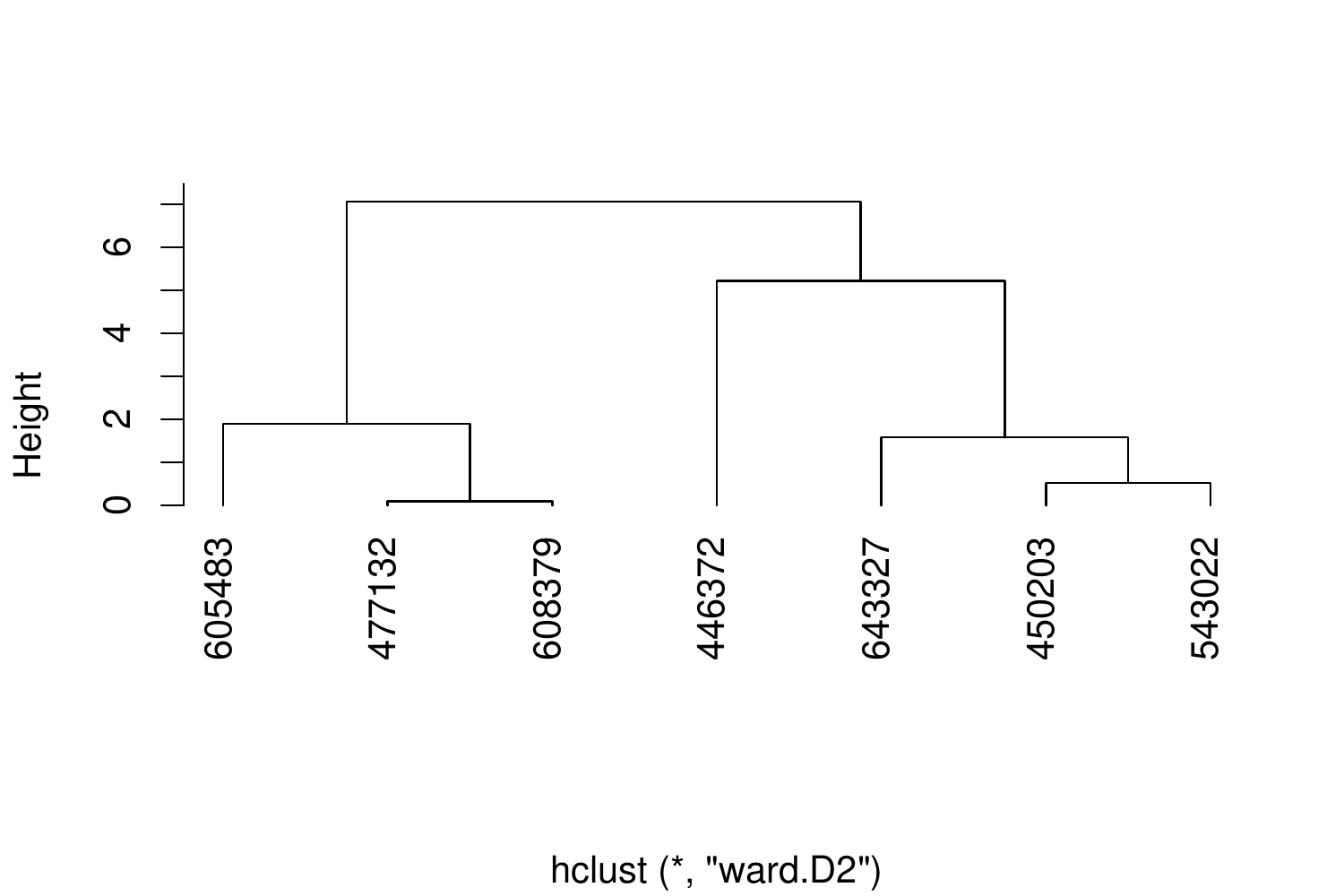}
			\includegraphics[width=0.25\textwidth]{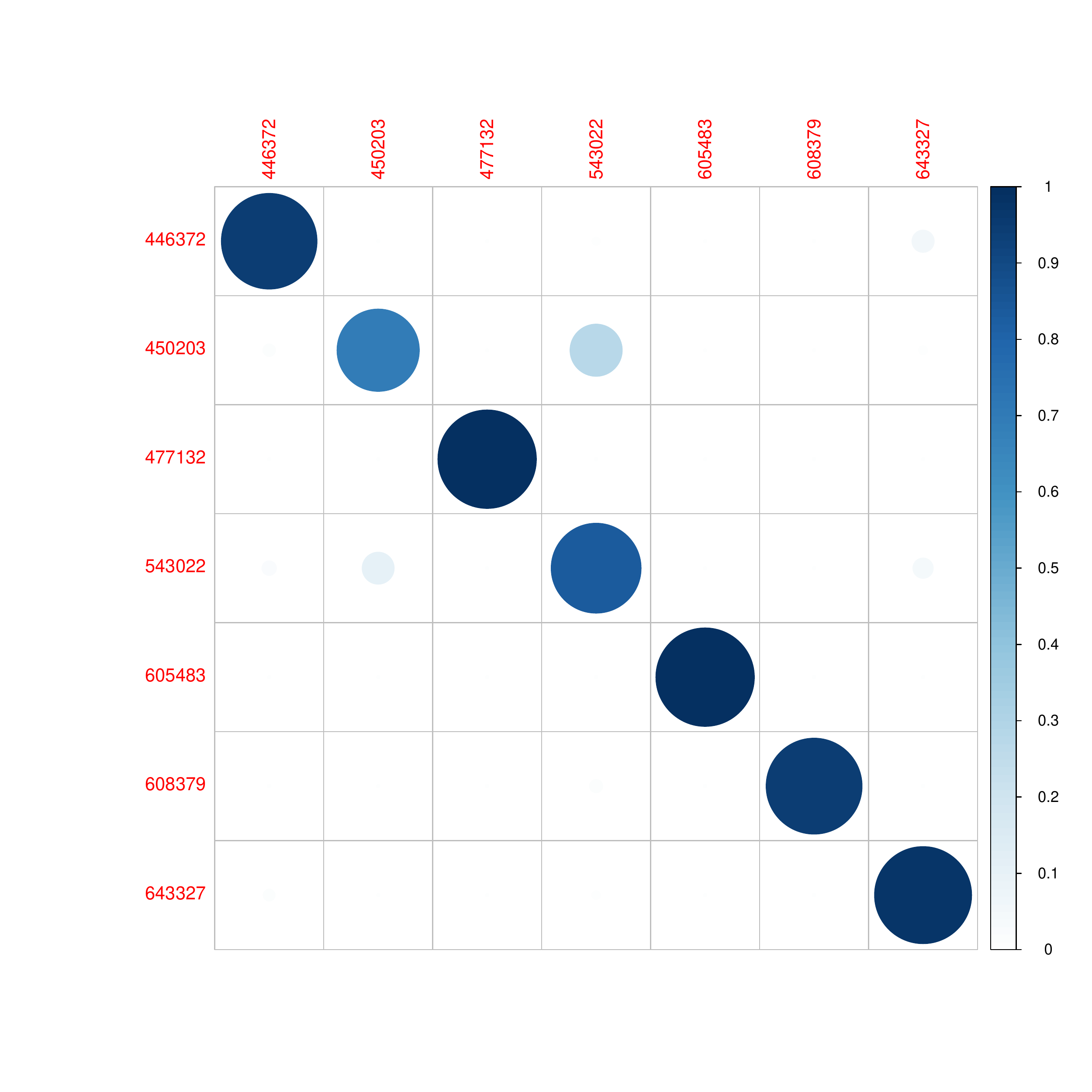}
			\caption{Curveball's Label embedding tree of feature set DEF and predictive graphs and predictive matrix (with threshold value 1) with respect to feature-sets: DEF.}
			\label{fig:curveball}
		\end{figure*}

With the same 4-to-1 training-testing ratio, our CEDA follows the same two routes as we previously have done under Slider's MCC setting. Our chain of feature-set begins with a feature-set, denoted as DEF, including members from three feature-groups: D, E, and F, marked in panel-B of Fig. \ref {fig:MCEM}. The LET of feature-set DEF and its predictive map with threshold being equal to 1 are given in panels-A\&B of Fig. \ref {fig:curveball}, respectively. Such a predictive map again shows asymmetry of mixing geometries, and the amounts of errors by forcefully committing to singleton predicted label.
		
In contrast, with the choice of threshold range $[C_L, C_U]=[0.65, 100/65]$ for PL-ratio, the resultant predictive map of feature-set DEF is reported in Table~\ref{tab:curveDEF}. This feature-set DEF seems rather efficient. We see that, except pitcher-b (with ID:450203), the rest of 6 pitchers all are nearly exclusively predicted. Pitcher-IDs are encoded as following: pitcher-a(446372); pitcher-c(477132); pitcher-d(543022); pitcher-e(605483); pitcher-f(608379) and pitcher-g(643327). Among the 12 mixing geometric pattern categories of 1st order, 6 are pitcher-ID specific, while 6 categories are having uncertainty. Next, we choose the feature set BB to be the 2nd feature-set in our chain of feature-sets to further explore the fine scale MCC information content of curveball.
		
		\begin{table*}[!htb]
			\caption{The predictive map matrix from feature-set DEF. (*) marks categories with uncertainty.}
			\label{tab:curveDEF}
			\begin{tabular}{|l|lllllll|}\hline
				DEF    & a   & b   & c  & d  & e  & f  & g   \\\hline
				a      & 146 & 0   & 0  & 0  & 0  & 0  & 1   \\
				*b     & 0   & 10  & 0  & 2  & 0  & 0  & 0   \\
				c      & 0   & 0   & 80 & 0  & 0  & 0  & 0   \\
				*d     & 0   & 4   & 0  & 30 & 0  & 1  & 0   \\
				e      & 0   & 0   & 0  & 0  & 44 & 0  & 0   \\
				f      & 0   & 0   & 0  & 0  & 0  & 54 & 0   \\
				*g     & 1   & 0   & 0  & 0  & 0  & 0  & 135 \\
				cef    & 0   & 0   & 2  & 0  & 0  & 0  & 0   \\
				*bd    & 0   & 111 & 0  & 10 & 0  & 0  & 0   \\
				*bdg   & 0   & 1   & 0  & 5  & 0  & 0  & 4   \\
				*abdg  & 13  & 3   & 0  & 1  & 0  & 0  & 4   \\
				all    & 0   & 0   & 0  & 0  & 0  & 4  & 0  \\\hline
			\end{tabular}
		\end{table*}

With the chain of 2 feature-sets: DEF to BB, the results of such chain is reported in Table~\ref{tab:curveDEFBB}. Majority of 2nd order mixing geometric pattern categories turns out to be pitcher specific, while those having uncertainty have relative extreme odds-ratios. We surely can explore further. We stop here to avoid repeating similar messages on the effectiveness of such chains of feature-sets. The above two tables of local mixing geometric pattern categories demonstrates effects of two complementary feature-sets, and at the same time jointly reveal the curveball's MCC information content. Such a chain of 2 feature-sets clearly offers perfect predictive decision-making with explainable supporting evidences. Likewise, the platform consisting the 1st and 2nd orders of mixing geometric pattern categories can map out the uncertainty of machine learning methodologies, as would be discussed below.
		
		\begin{table*}[!htb]
			\caption{Classification results of chain of two feature-sets: DEF to BB. }
			\label{tab:curveDEFBB}
			\begin{tabular}{|l|llll||l|llll||l|llll|}\hline
				*b   & a & b & d & g & *d    & a  & b & d  & g & *g  & a & b & d & g  \\\hline
				b    & 0 & 5 & 1 & 0 & d     & 0  & 2 & 20 & 0 & a   & 1 & 0 & 0 & 0  \\
				abdf & 0 & 1 & 0 & 0 & f     & 0  & 0 & 2  & 0 & f   & 0 & 0 & 0 & 1  \\
				bd   & 0 & 4 & 1 & 0 & abdf  & 0  & 2 & 0  & 0 & g   & 0 & 0 & 0 & 91 \\
				     &   &   &   &   & bd    & 0  & 0 & 2  & 0 & bdf & 0 & 0 & 0 & 2  \\
				     &   &   &   &   & bdf   & 0  & 0 & 4  & 0 & all & 0 & 0 & 0 & 41 \\
				     &   &   &   &   & all   & 0  & 0 & 2  & 0 &     &   &   &   &    \\
				\hline\hline
				*bdg & a & b & d & g & *abdg & a  & b & d  & g & *bd  & a & b & d & g \\\hline
				b    & 0 & 1 & 0 & 0 & a     & 11 & 0 & 0  & 0 & b &0 & 12 & 1 & 0\\
				d    & 0 & 0 & 3 & 0 & g     & 0  & 0 & 0  & 3 &d  & 0 & 2 & 4 &0\\
				g    & 0 & 0 & 0 & 1 & abdf  & 2  & 2 & 1  & 0 & abdf&0&8&0&0   \\
				bd   & 0 & 0 & 2 & 0 & bd    & 0  & 1 & 0  & 0 &  bd&0&89&5&0 \\
				bdf  & 0 & 0 & 0 & 1 & bdf   & 0  & 0 & 0  & 1 &  &&&&   \\
				all  & 0 & 0 & 0 & 2 &       &    &   &    &   &    &&&&  \\ \hline
				
			\end{tabular}
		\end{table*}

\subsection{Dissecting uncertainty of results from machine learning algorithm}	
Under MCC settings, machine learning (ML) algorithms, like random forest \cite{breiman01} and various boosting approaches \cite{freund97,zhu09}, are popularly employed. They are able to achieve low classification error rates. Such successes are particularly commonly seen when the number of features $K$ is large relative to the size of label space $L$. One factor certainly has been contributing significantly to such successes. This factor is that the complexity of unseen decision trees is never revealed and seldom concerned in such applications in literature. Having a large $K$, all involving decision-trees can grow very tall and spread very wide. The cost of such ``model'' complexity has been left out of considerations when making decisions. All these decisions are derived from black-boxes in the sense of no reasons and interpretations attached. This issue of non-interpretability is serious on top of the complexity issue. Therefore, even potentially having low error rates, all these machine learning results under MCC settings are subject to uncertainty of being simultaneously right and wrong.

Our MCC information content indeed can serve as an ultimate standard of validity for uncertainty resulted from machine learning algorithms. The validity check is simply subject each testing data point to both CEDA and any machine learning algorithm. Then we view and check which one of mixing geometric pattern categories a testing data point's machine learning (ML) decision falls into. Such a check leads one of the following 4 possibilities: a ML decision lands in mixing geometric pattern category
\begin{description}
\item[1.[certainty-coherent]]: with certainty, and they are coherent;
\item[2.[certainty-incoherent]]: with certainty, but they are incoherent;
\item[3.[uncertainty-coherent]]:with uncertainty and they are coherent;
\item[4.[uncertainty-incoherent]]:with uncertainty, but they are incoherent;
\end{description}
The [certainty-coherent] case serves a confirmation of the ML-decision. Both [certainty-incoherent] and [uncertainty-incoherent] are likely pointing to that the ML' predictive result is``definitely'' wrong. There exists still uncertainty when for a ML predictive result falling into [uncertainty-coherent] case. However, it will be 100\% correct if we simply report our mixing geometric pattern category as a decision-set. Upon these four cases, all ML methodologies seem evidently can be significantly strengthened by projecting their predictive results with respect to our MCC information content. At the same time, we resolve the interpretation issue completely.

We illustrate such conclusions through applying Random Forest on slider and curveball data sets. In our experiments, indeed we have applied random forest multiple times and two aforementioned versions of boosting multiple times as well. All results are rather consistent with our conclusions presented here. So we only report results from random forest.

In the Table~\ref{tab:sliderRF}, we see that Random Forest (RF) makes 13 errors: 2 from pitcher-b being assigned to [pitcher-e], 1 from pitcher-c being assigned to [pitcher-b] and 10 from pitcher-e being assigned to [pitcher-b]. With respect to the 1st order mixing geometric pattern categories, the [certainty-incoherent] error of [pitcher-b] prediction on one pitch of pitcher-c has been confirmed at the category [c]. Among 12 remaining errors going into the 2nd order mixing geometric pattern categories in Table~\ref{tab:tab2}, we see that 1 RF's error is confirmed in category [b-e] in Table~\ref{tab:RF1st}. The 11 remaining errors going into the third order mixing geometric pattern categories in Table~\ref{tab:tab4}, we see 2 pitches of pitcher-e are confirmed in [e-e-be] and [e-e-abe], the rest of 9 pitches are in [uncertainty-coherent] case: 2 from pitcher-b and 7 from pitcher-e, are all in mixing geometric pattern categories with uncertainty.

That is, such a dissection on uncertainty of errors from Random Forest brings out the fact that all its errors are likely coming from rather intensive mixing localities of geometry of all involving point-clouds. From this perspective, it would be advantageous to make ML predictions with projections into mixing geometric pattern categories of MCC information content. This conclusion is true throughout all experiments via two versions of boosting as well.

\begin{table}[!htb]
			\caption{Results of one random forest application on slider data with errors indicated.}
			\label{tab:sliderRF}
			\begin{tabular}{|l|lllll|}\hline
				RF   & a   & b   & c  & d  & e \\\hline
				a    & 124 & 0   & 0  & 0  & 0  \\
				b    & 0   & 85  & 1  & 0  & 10  \\
				c    & 0   & 0   & 90 & 0  & 0  \\
				d    & 0   & 0   & 0  & 56 & 0  \\
				e    & 0   & 2   & 0  & 0  & 35 \\\hline
			\end{tabular}
	\end{table}

\begin{table}[!htb]
			\caption{Locations of random forest's errors on slider w.r.t the: 1st order predictive map (Left); 2nd order predictive map (Middle); 3rd order predictive map (Right). }
			\label{tab:RF1st}
			\begin{tabular}{|l|lll||l|ll||l|ll|}\hline
				RF     & b   & c   & e &RF        & b     & e& RF        & b     & e \\\hline
				b      & 0   & 0   & 1 &b-e       & 0     & 1&e-be-be   & 2     & 0 \\
				c      & 0   & 1   & 0 &e-e       & 0     & 2&be-b-be   & 0     & 1 \\
				e      & 2   & 0   & 2 &e-be      & 2     & 0&be-b-all  & 0     & 1 \\
				be     & 0   & 0   & 6 &be-b      & 0     & 2&be-be-be  & 0     & 3 \\
				all    & 0   & 0   & 1 &be-be     & 0     & 4&be-be-all & 0     & 1 \\
				&&&&all-be    & 0     & 1 &all-be-all& 0     & 1 \\\hline
			\end{tabular}
		\end{table}

To iterate such a conclusive statement, we again view the errors made by Random Forest on curveball data set from the perspective of its MCC information content. There are 14 errors committed by Random Forest on one application, as shown in Table~\ref{tab:curveRF}. Upon Table~\ref{tab:RFC1st}, all 14 errors are falling into 1st order mixing geometric pattern categories with uncertainty as shown in Table~\ref{tab:curveDEF}. Further, upon Table~\ref{tab:RFC1st} and due to one (d-bd) and one (bdg-db) being corrected,  12 out of 14 fall into the 2nd order mixing geometric pattern categories as shown in Table~\ref{tab:curveDEFBB}. Likewise, majority of errors from two version of boosting approaches is also found in a MCC's mixing geometric pattern category with uncertainty. Again we iterate that it would be advantageous to make ML predictions with projections into mixing geometric pattern categories of MCC information content.

\begin{table*}[!htb]
			\caption{Results of one random forest application on curveball data with errors indicated.}
			\label{tab:curveRF}
			\begin{tabular}{|l|lllllll|}\hline
				RF    & a   & b   & c  & d  & e  & f  & g   \\\hline
				a    & 158 & 0   & 0  & 0  & 0  & 0  & 1   \\
				b    & 2   & 128 & 0  & 9  & 0  & 0  & 0   \\
				c    & 0   & 0   & 82 & 0  & 0  & 0  & 0   \\
				d    & 0   & 1   & 0  & 37 & 0  & 0  & 0   \\
				e    & 0   & 0   & 0  & 0  & 44 & 0  & 0   \\
				f    & 0   & 0   & 0  & 0  & 0  & 59 & 0   \\
				g    & 0   & 0   & 0  & 2  & 0  & 0  & 144 \\\hline
			\end{tabular}
	\end{table*}

\begin{table*}[!htb]
			\caption{Locations of random forest's errors on curveball w.r.t the: 1st order predictive map (Left); 2nd order predictive map (Right).}
			\label{tab:RFC1st}
			\begin{tabular}{|l|lll||l|lll|}\hline
				RF     & a   & b   & d &RF        & a   & b   & d \\\hline
				b      & 0   & 0   & 2 &b-b       & 0   & 0   & 1\\
				d      & 0   & 0   & 1 &b-bd      & 0   & 0   & 1\\
				bd     & 0   & 1   & 6&bd-b      & 0   & 0   & 1  \\
				bdg    & 0   & 0   & 1&bd-d      & 0   & 0   & 1  \\
				abdg   & 2   & 0   & 1&bd-bd    & 0   & 1   & 4 \\
				&&&&bdg-bd    & 0   & 0   & 1 \\
				&&&&abdg-abdf & 2   & 0   & 1 \\\hline
			\end{tabular}
		\end{table*}

\section{Response manifold analytics (RMA)}
Some features and feature-groups evidently don't shed informative lights on labels' idiosyncratic characteristics, such as \{``ax'', ``az'', ``pfx\textunderscore x'', ``pfx\textunderscore z'', ``break\textunderscore length'', ``break\textunderscore angle''\}. Together with ``end\textunderscore speed'', these features are all measured at the home-plate. They are response variables in nature. In contrast, the rest of 14 features, including those informative features for labels' characteristics, are all measured around the pitcher's mound. That is, they are covariate variables in nature. Thus, separations of response-vs-covariate variables are physical along the spatial axis with 60'-6" fts (18.39m) in length from the pitcher's mound to the home plate and along the temporal axis in a 0.5 (500ms) sec. span for a baseball to arrive home plate (w.r.t 90mph).

The collective associative relations of 7-response-to-14-covariate are contemplated as pitching mechanics as one whole, which is termed response manifold analytics (RMA). From physics perspective, these RMA associative relations are universally governed by Newton's laws of forces, which include gravity, and speed and pressure in aerodynamics and Magnus effects of spin and many bio-mechanical ones. The task of synthesizing these individually known forces into one single system of pitching dynamics is captured by ``Devils in the details''. That is, a functional system embracing all such intertwined and coupled forces in fine detail is neither known, nor existing. Nevertheless, we believe that PITCHf/x database likely captures such a functional system to a great extent, if not entirely. Thus, our challenge in the 2nd part of this paper is: {\bf Can we approximate such a functional system purely from Data Science perspective?}

To build a functional system that indeed conforms with all observed manifolds, we again adopt CEDA theme: Perform global-to-local and coarse-to-fine mutliscale explorations for response-to-covariate associative relations. Such CEDA computational endeavors are to identifying a set of major features and all involving minor features for each concerned response-to-covariate association. The chief merit of a set of major covariate features is that these features couple with response features to form an evident geometric manifold with low complexity. Further, upon such a manifold, the collection of rectangles or hypercubes framed by major features's categorical structures via their histograms induces a collection of localities. Each locality of a manifold is supposed to have rather simple and understandable geometric structures. Hopefully we can link such locality specific geometric structures into a well-established global approximation of the entire manifold.

In contrast, the merit of a minor feature rests on discovering its heterogeneous effect on a limited number of localities with respect to its categorical structures. This is the CEDA theme for computing global-to-locality information content from a manifold.

\begin{figure*}[h!]
\centering
\includegraphics[width=3.1in]{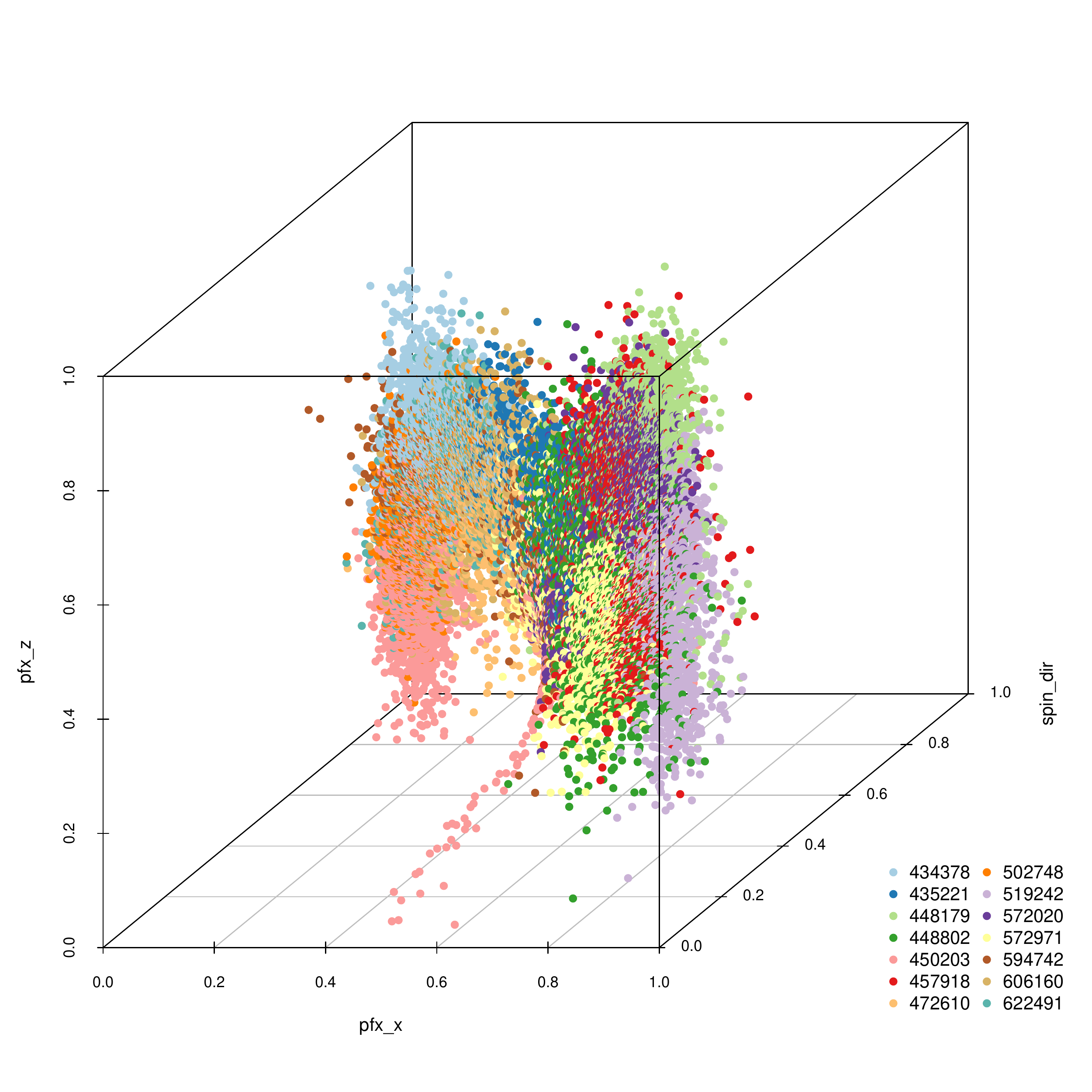}
\includegraphics[width=3.1in]{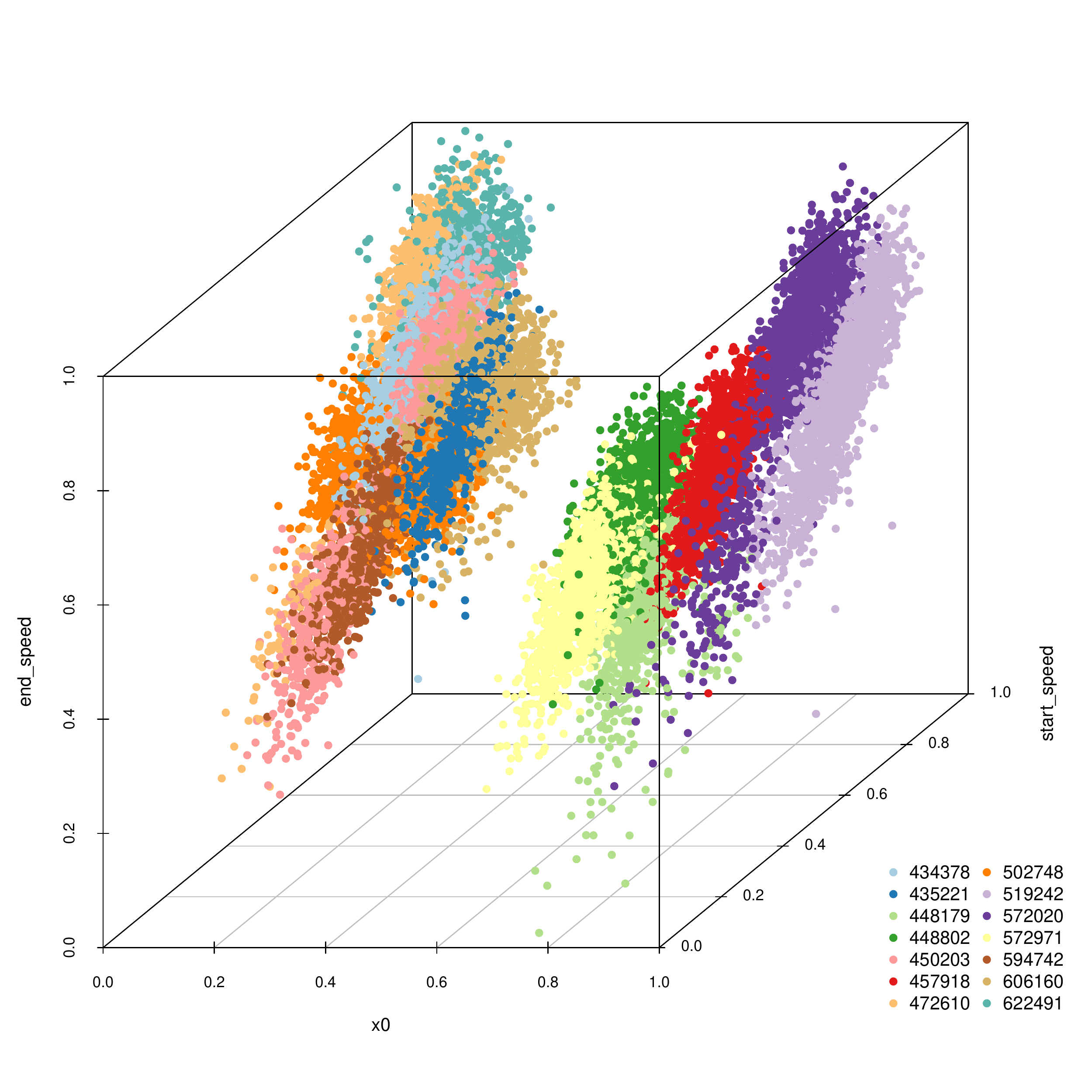}

\caption{Fastball's two manifolds: (A)\{``pfx\textunderscore x'', ``pfx\textunderscore z'', `` spin\textunderscore dir''\}; (B) \{``x0'', ``start\textunderscore speed'', ``end\textunderscore speed''\}, see corresponding rotatable 3D plots in (\url{https://rpubs.com/CEDA/baseball}).}
\label{fig:3Dxydir}
\end{figure*}

Here we develop fastball's RMA along the CEDA theme with focus placed on two intrinsically distinct manifolds, as shown in Fig. \ref{fig:3Dxydir}, that are primary mechanisms of pitching dynamics of fastball. Likewise developments can be done for slider and curveball's RMA.

\subsection{Manifolds on \{``pfx\textunderscore x'', ``pfx\textunderscore z'', `` spin\textunderscore dir''\}}
Even from hindsight, we know that Magnus effect is indeed underlying the 2D manifolds of \{``pfx\textunderscore x'', ``pfx\textunderscore z'', `` spin\textunderscore dir''\} in $R^3$ as vividly demonstrated through rotatable 3D plots: slider and fastball (\url{https://rpubs.com/CEDA/baseball}) with their snapshots shown in panel-D of Fig. \ref{fig:infogeoS} and panels-A of Fig. \ref{fig:3Dxydir}, respectively. The two manifolds show strikingly explicit differences in geometric characteristics pertaining to pitch types. Mechanically speaking, it means that Magnus effects operates in distinct manners within distinct domains of feature values with respect to the two pitch-types. Mathematically, the two systems of functions describing these 2D manifolds are highly nonlinear across the entire span of pitching dynamics. Hence it would be reasonable to explore a pitch-type's functional systems one at a time.

\begin{figure}[h!]
\centering
\includegraphics[width=3.1in]{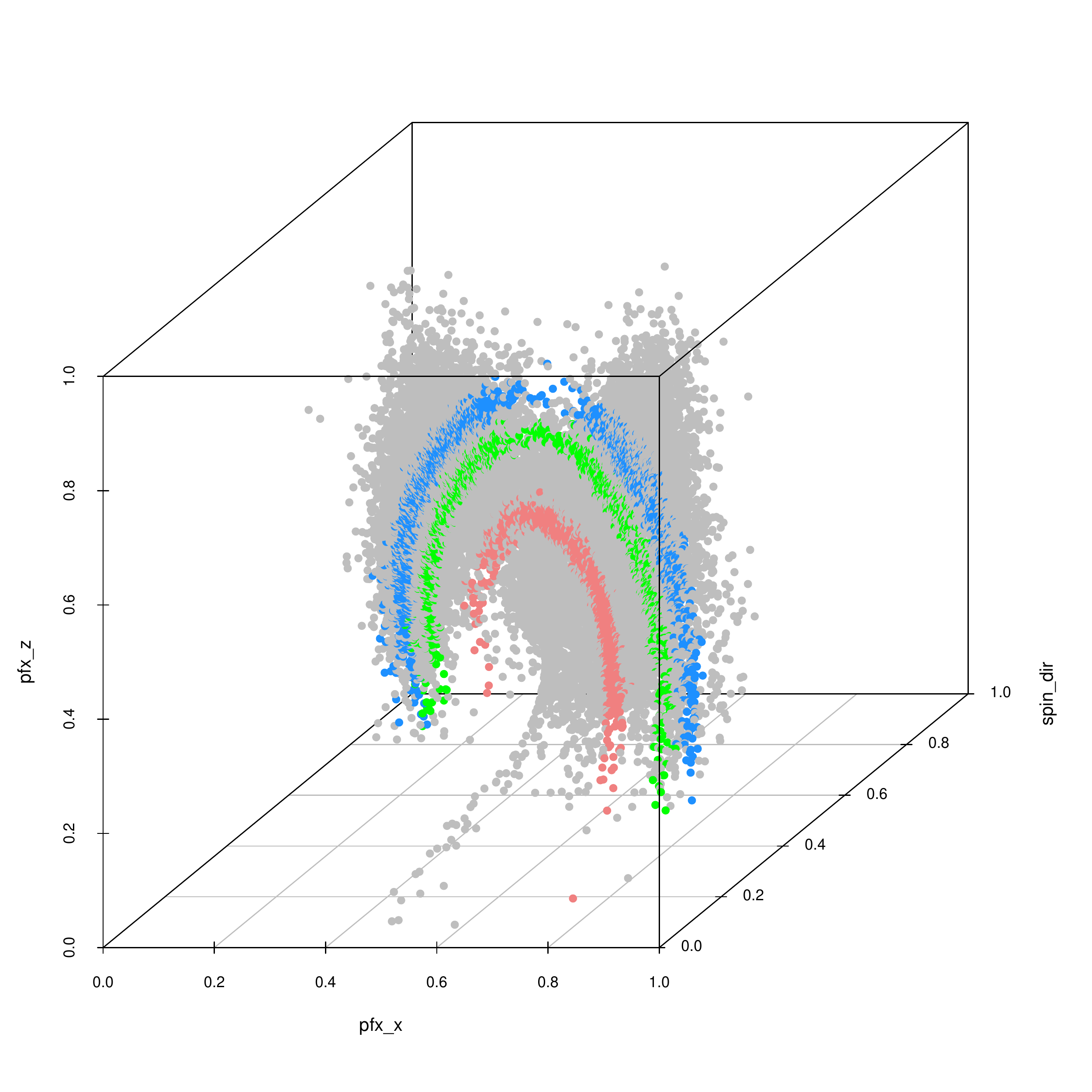}
\includegraphics[width=3.1in]{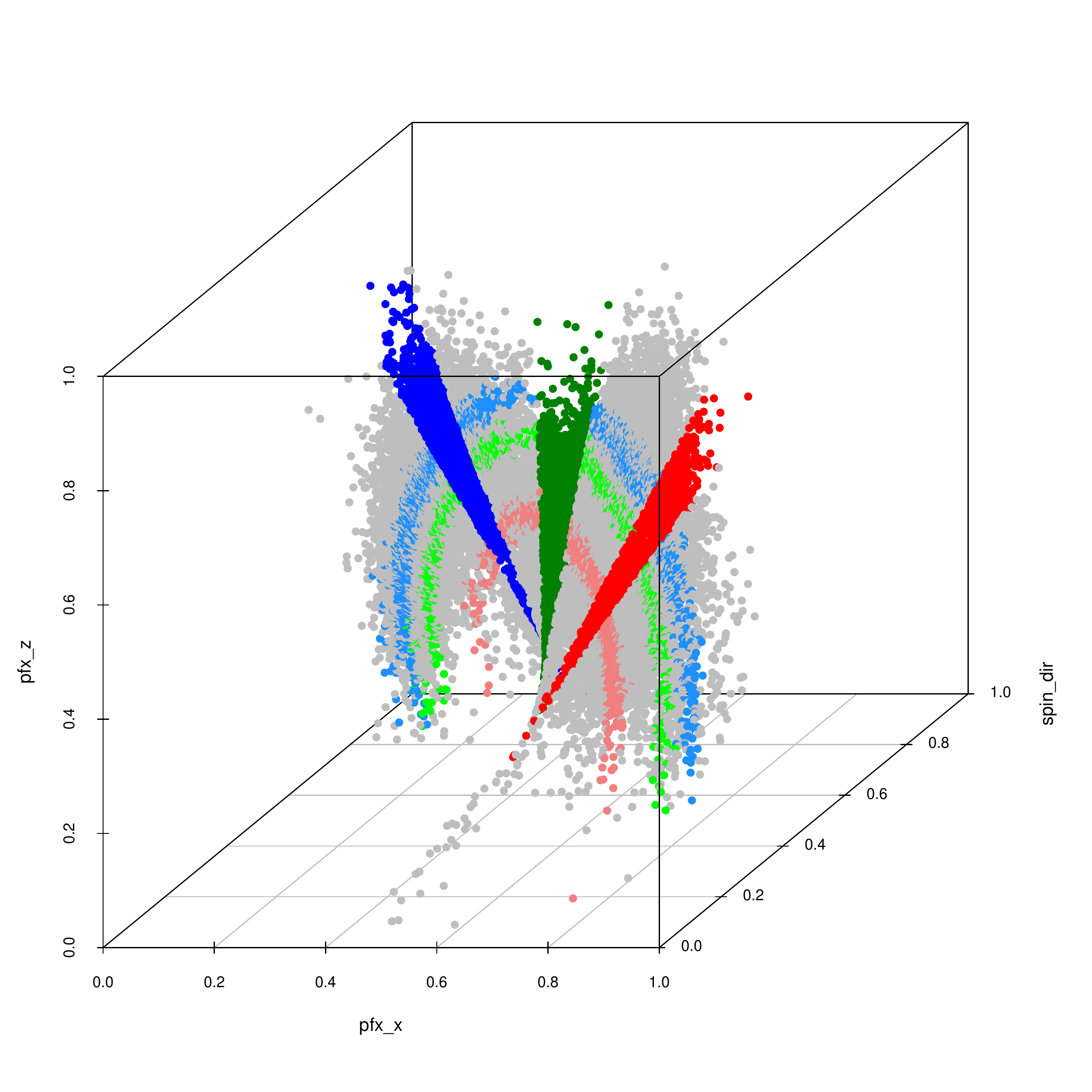}

\caption{Manifolds of \{``pfx\textunderscore x'', ``pfx\textunderscore z'', ``spin\textunderscore dir''\} with marked strips: (A) 3 strips of ``spin\textunderscore rate''; (B) 3 strips of ``spin\textunderscore rate'' intersecting with 3 strips ``spin\textunderscore dir''. See corresponding rotatable 3D plots in (\url{https://rpubs.com/CEDA/baseball}).}
\label{fig:3and6striprate}
\end{figure}

\begin{figure}[h!]
\centering
\includegraphics[width=3.1in]{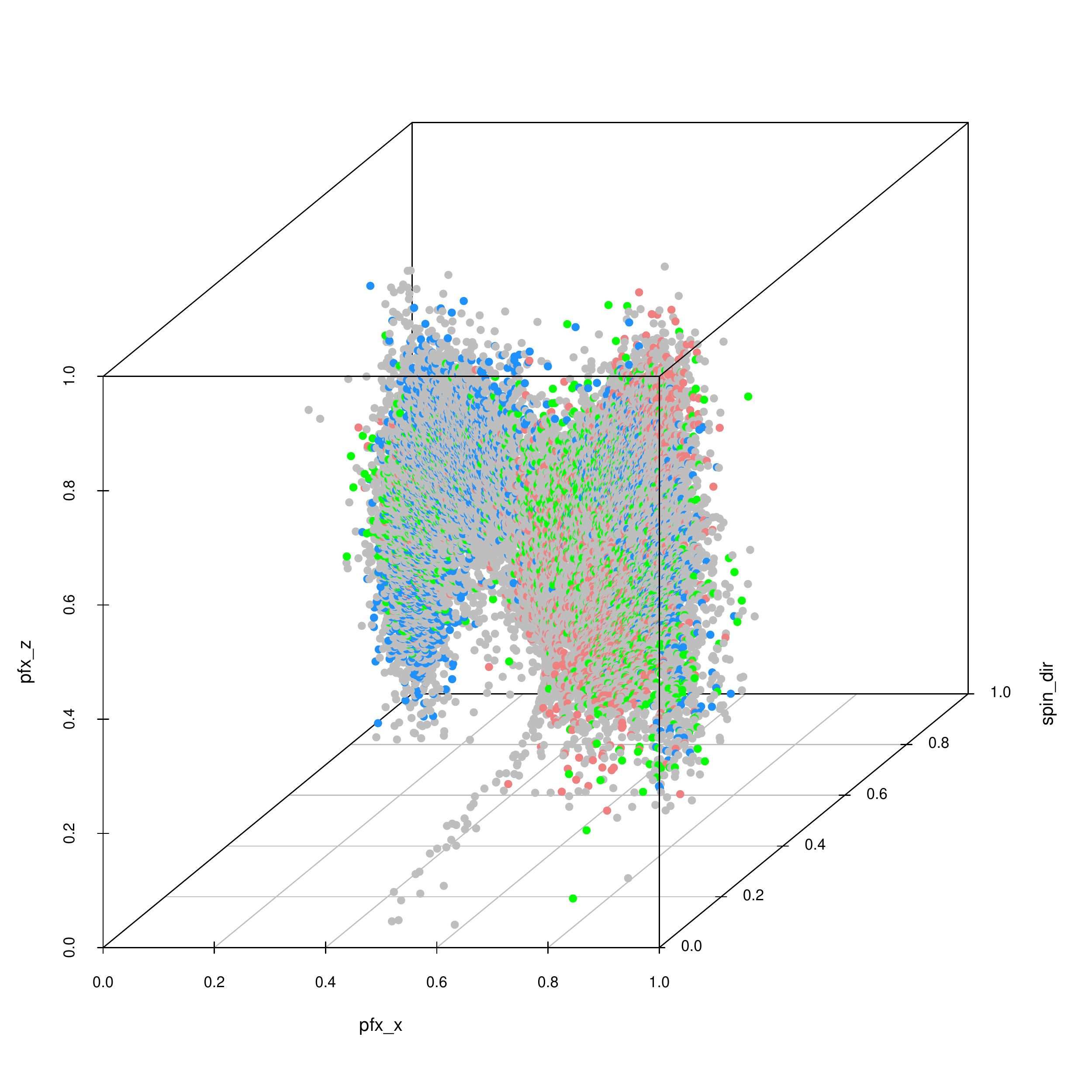}
\includegraphics[width=3.1in]{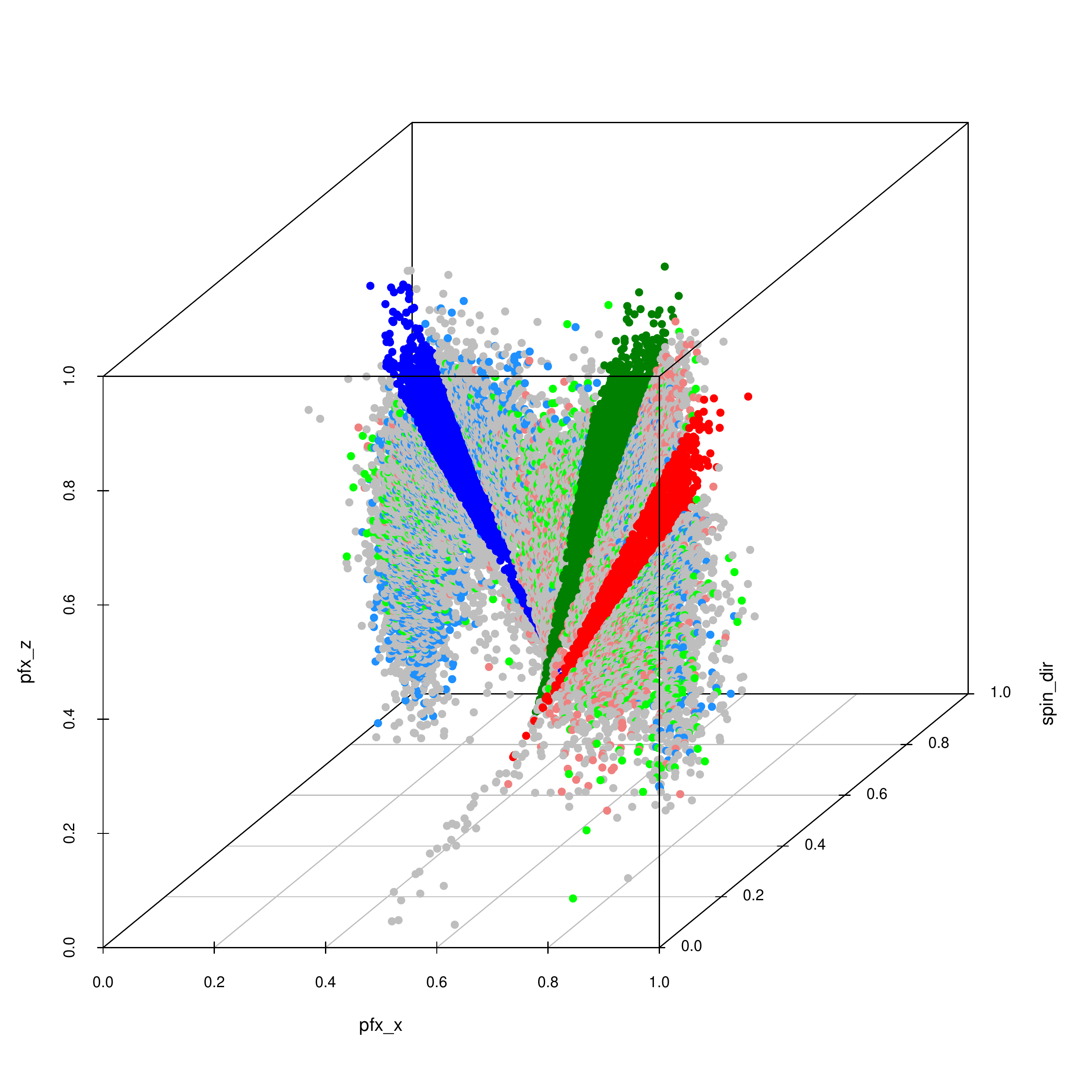}

\caption{Manifolds of \{``pfx\textunderscore x'', ``pfx\textunderscore z'', ``spin\textunderscore dir''\} with marked strips: (A) 3 strips of ``start\textunderscore speed''; (B) 3 strips of ``start\textunderscore speed'' intersecting with 3 strips ``spin\textunderscore dir''. See corresponding rotatable 3D plots in (\url{https://rpubs.com/CEDA/baseball}).}
\label{fig:3and6stripspeed}
\end{figure}

For expositional simplicity, our computational developments for RMA in this section are illustrated through fastball. We begin fastball's RMA by explicitly exploring major features determining response features \{``pfx\textunderscore x'', ``pfx\textunderscore z''\} among 14 covariate features plus the categorical one of pitch-IDs. Upon the discovery of 2D manifolds of \{``pfx\textunderscore x'', ``pfx\textunderscore z'', `` spin\textunderscore dir''\} in $R^3$, as shown in of panel-A Fig. \ref{fig:3Dxydir}, we know that `` spin\textunderscore dir'' is one major feature. We seek for other major feature by color-coding the candidate feature's bins, and then marking them onto the manifold of \{``pfx\textunderscore x'', ``pfx\textunderscore z'', `` spin\textunderscore dir''\}. Each of a major feature's bins will mark a clear-cut strip on this manifold. For example, the bins of `` spin\textunderscore rate'' mark clear-cut strips, as shown in panel-A of Fig. \ref{fig:3and6striprate}. But it is not the case for `` start\textunderscore speed'', as shown in panel-A of Fig. \ref{fig:3and6stripspeed}. No clear-cut strips in sight, but defused colored points all over the manifold. This defusing pattern indicates that`` start\textunderscore speed'' is not a major feature of this manifold.

If we also mark 3 strip of `` spin\textunderscore dir'' from 3 separate bins of its histogram, we see clearly 9 more or less rectangle intersecting patches, as shown in panel-B of Fig. \ref{fig:3and6striprate}. Therefore, all bins of the histogram of `` spin\textunderscore dir'' coupled with all bins of the histogram of `` spin\textunderscore rate'' frame a lattice of patches.  And each patch in this patching lattice is a locality of \{`` spin\textunderscore dir'',`` spin\textunderscore rate'',``pfx\textunderscore x'', ``pfx\textunderscore z''\} 4D manifold. After discovering a set of major feature and constructing a patching ensemble, we mount to be able to approximate the entire manifold as one whole. This approximation scheme is seen as one of the chief merits of using a set of major covariate feature to understand how targeted response features come to be as they are. It also becomes clear that the defusing patterns induced by including a non-major feature, such as `` start\textunderscore speed'', as seen in Fig. \ref{fig:3and6stripspeed}, can't afford reasonable approximations across the entire range of a targeted response features.

\begin{figure}[h!]
\centering
\includegraphics[width=3.1in]{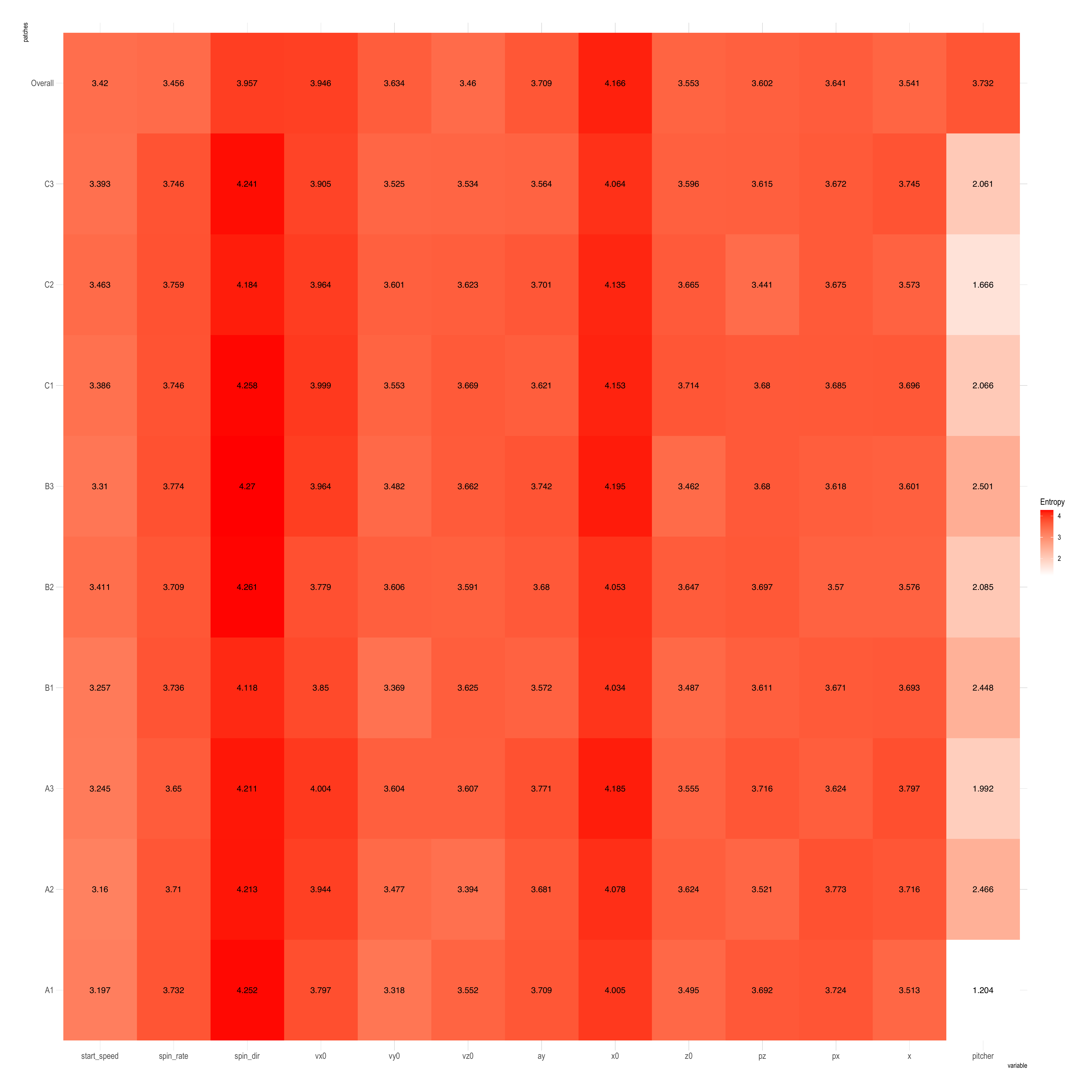}

\caption{Minor features selection based Shannon entropy on 9 patches on manifolds of \{``pfx\textunderscore x'', ``pfx\textunderscore z'', ``spin\textunderscore dir''\} framed by 9  intersecting patches via 3 strips of ``spin\textunderscore dir'' with 3 strips ``spin\textunderscore dir'' as seen in Fig. \ref{fig:3and6striprate}.}
\label{fig:minor9}
\end{figure}

After discovering the \{``spin\textunderscore dir'', ``spin\textunderscore rate''\} being a set of major features to the target response features \{``pfx\textunderscore x'', ``pfx\textunderscore z''\}, the resultant lattice of localities of 4D manifold becomes the basis for exploring all potential minor features. It is worth reiterating that a minor feature could reveal its different effects on some localities, while having no effects at all on the rest of localities. One illustrative example feature is the pitcher-ID. As revealed in panel-A of Fig. \ref{fig:3Dxydir}, some pitcher-IDs are dominant on some localities, some are intensively mixed on some other localities. Therefore it is essential to figure out and document which features' which bins are dominant on which localities. Shannon entropy can naturally bring out degrees of such kind of dominance of bin-memberships. We report the Shannon entropies with respect to the 9 patches identified in panel-B of Fig. \ref{fig:3and6striprate} across all covariate features and Pitcher-ID in the Fig. \ref{fig:minor9}. We see that the pitch-ID is the obvious minor feature. Such finding is basically categorical.

The effort of identifying all potential minor features within each locality of 4D manifold \{``spin\textunderscore dir'', ``spin\textunderscore rate'', ``pfx\textunderscore x'', ``pfx\textunderscore z''\} is crucial for seeing the spectrum of information content contained in PITCHf/x about pitching dynamics of fastball. Collectively all minor features' heterogeneous effects across all localities become a map of heterogeneity pertaining to all relevant minor features. This map helps major features to provide very detailed information about MLB pitching dynamics from the perspective of response features \{``pfx\textunderscore x'', ``pfx\textunderscore z''\}.

In summary, on the task of figuring out how Magnus effects works out fastball's \{``pfx\textunderscore x'', ``pfx\textunderscore z''\}, the discovered major features \{``spin\textunderscore dir'', ``spin\textunderscore rate''\} give us a global view as well as local view through the 4D manifold. The identified minor features reveal their roles within localities of this manifold. This major-to-minor is meant for global-to-fine scales of information content. This theme echoes Nobel laureate physicist P. W. Anderson's statements in his 1972 paper with title ``More is different'': the twin difficulties of scale and complexity (heterogeneity) are keys to understanding large complex systems.

We also discover two more manifolds: \{``spin\textunderscore dir'', ``ax'', ``az''\} and \{``break\textunderscore length'', ``break\textunderscore angle'',``spin\textunderscore dir''\}. It is not surprising that both manifold are very similar with the manifold \{``pfx\textunderscore x'', ``pfx\textunderscore z'', `` spin\textunderscore dir''\} because, as shown in the mutual conditional entropy matrix in Fig. \ref{fig:MCEM}, \{``spin\textunderscore dir'', ``ax'', ``pfx\textunderscore x'', ``break\textunderscore angle''\} form a tightly associative feature group, so do \{``spin\textunderscore rate'', ``az'', ``pfx\textunderscore z'', ``break\textunderscore length''\}. Therefore, we don't need to repeat the same RMA for response features \{``ax'', ``az'', ``break\textunderscore length'', ``break\textunderscore angle''\}.

The above computational results of RMA imply that our predictive decision-making for \{``pfx\textunderscore x'', ``pfx\textunderscore z''\}, \{``ax'', ``az''\} or \{``break\textunderscore length'', ``break\textunderscore angle''\} must be performed following this {\bf inferential protocol}: 1) first, identify a proper locality framed by the major features; 2) choose a proper and focal neighborhood system within this locality, such as $K$-nearest neighbors; 3) use each identified minor feature as a sieve to filter out incoherent neighbors. We are confident that this intrinsic global-to-local idea underlying this RMA-based prediction protocol is universal for all complex systems. We discuss the performance of such a RMA based inferential protocol in the next subsection.

\subsection{Predictive inferences via a manifold on \{``pfx\textunderscore x'', ``pfx\textunderscore z'',\\ `` spin\textunderscore dir''\}}
After identifying major and minor feature with respect to a set of multiple response features, it is natural and important to discuss the predictive inference as the next computational task. Since this discussion will further reveal merits of our CEDA's global-to-local theme. That is, we want to clearly demonstrate impacts of manifold on predictive decision-making. Hopefully this message can bear essential impacts on any other databases beyond PITCHf/x. In addition, we hope such RMA based inferential protocol would further echo physicist P. W. Anderson's universal view on large complex systems through Data Science.

Let a target set of response variables be \{``pfx\textunderscore x'', ``pfx\textunderscore z''\}, denoted as $\{Y_1, Y_2\}$. Upon 4D manifold of \{``pfx\textunderscore x'', ``pfx\textunderscore z'', ``spin\textunderscore dir'', ``spin\textunderscore rate''\}, we have identified a set of major features: \{`` spin\textunderscore dir'', `` spin\textunderscore rate''\}, denoted as $\{X_1, X_2\}$, and a minor feature ``pitcher-ID'', denoted as $Z_1$. How to predict $\{Y^*_1, Y^*_2\}$ given $\{x_1, x_2, z_1\}$.

Since features ``pfx\textunderscore x'' and ``pfx\textunderscore z'' are evidently associated with each other. A reasonable predictive decision-making mechanism is to estimate such a bivariate response variable $(Y^*_1, Y^*_2)$ simultaneously. According the aforementioned RMA-based inferential protocol, our predictive decision-making process is given as follows.

Identify a $2$-dim training-covariate-rectangle ${\cal B}<x_1, x_2>=B[X_1]\times B[X_2]$, where $x_1 \in B[X_1]$ and $x_2 \in B[X_2]$ with $B[X_1]$ and $B[X_2]$ being one bin of the two histograms of features $X_k$ with $k=1, 2$ based on training data, respectively. Also we denote the training-response-region of ${\cal B}<x_1, x_2>$:
\[
{\cal F}[{\cal B}<x_1, x_2>]=\{(Y_1, Y_2)| (X_1, X_2)\in {\cal B}<x_1, x_2>\}.
\]
That is, ${\cal F}[{\cal B}<x_1, x_2>]\times {\cal B}<x_1, x_2>$ is a patch (so-called locality) shown in Fig. \ref{fig:3and6striprate} on the manifold \{``pfx\textunderscore x'', ``pfx\textunderscore z'', `` spin\textunderscore dir'',`` spin\textunderscore rate''\}. This locality is the proper region of interest that captures the global scale of pattern information upon the manifold on an identified local covariate region of ${\cal B}<x_1, x_2>$. Our predictive decision-making on $(Y^*_1, Y^*_2)$ is proceeded by applying the $k^*$-nearest neighbor(KNN) approach with $k^*=20$. Let the set of $k^*$ nearest neighbors be denoted by $B<x_1, x_2>=\{(X_1, X_2)^{h}\in {\cal B}|h=1, 2,...,20\}$. Next we define $B^{z_1}<x_1, x_2>$ as a subset containing members of $B<x_1, x_2>$ that are actually associating with the observed minor feature $z_1$.

Then we define a focal subset of ${\cal F}[{\cal B}<x_1, x_2>]$ given $(x_1, x_2, z_1)$as:
\[
{\cal F}[B^{z_1}<x_1, x_2>]=\{(Y_1, Y_2)| (X_1, X_2)\in B^{z_1}<x_1, x_2>\}.
\]
Then we make a predicted vector $(\hat{Y}^*_1, \hat{Y}^*_2)$ as the center of mass, or simple average, of focal subset ${\cal F}[B^{z_1}<x_1, x_2>]$.

In this manner, this estimation of $(\hat{Y}^*_1, \hat{Y}^*_2)$ indeed incorporate global and local structures of major and minor features. We see this predictive decision-making simultaneously embraces geometric patterns of manifold \{``pfx\textunderscore x'', ``pfx\textunderscore z'', `` spin\textunderscore dir'',`` spin\textunderscore rate''\} that capture the multiscale associations between bivariate response s\{``pfx\textunderscore x'', ``pfx\textunderscore z''\}  and bivariate major covariate \{`` spin\textunderscore dir'', `` spin\textunderscore rate''\} and takes care of heterogeneity due to minor feature ``pitcher-ID'' within locality.

Here we perform an experiment by comparing our RMA based predictive decision-making with Random Forest regression. Here Random Forest regression is applied to estimate $Y^*_1$ and $Y^*_2$, separately with respect to all covariate 15 features. The values of tuning parameters used in Random Forest regression are all default choices. We consider three kinds of sum of errors: 1) individual sum of squared errors for $Y^*_1$(``pfx\textunderscore x'') and $Y^*_2$(``pfx\textunderscore z''), see Table~\ref{tab:SSEINV}; 2) bivariate correlated sum of squared error with respect to overall training covariance matrix of $\{Y_1, Y_2\}$, see Table~\ref{tab:BSSEALL}; 3) bivariate correlated sum of squared error with respect to ${\cal B}<x_1, x_2>$ locality-specific training covariance matrix of $\{Y_1, Y_2\}$, see Table~\ref{tab:BSSELOCAL}. Here the bivariate correlated sum of squared errors is calculated by re-scaling the bivariate error vector with respect to its corresponding covariance matrix, which is either based on all training data of $\{Y_1, Y_2\}$, or locality-specific $\{Y_1, Y_2\}$ corresponding to identified rectangles ${\cal B}<x_1, x_2>$.

Across these three tables, our RMA-based predictive decision-making overall significantly out perform Random Forest Regression on 8 of 9 patches marked in Fig. \ref{fig:3and6striprate} on the manifold \{``pfx\textunderscore x'', ``pfx\textunderscore z'', `` spin\textunderscore dir''\} in term of individual response features as well as bivariate vector. Both approaches are rather equal in the C1 patch. The bivariate correlated sum of squared error with respect to overall training covariance matrix is particularly outstanding, upon which we see merits on RMA-based inferential protocol. It is noted that patch-wise covariance matrix re-scaling might be too volatile as seen in Table~\ref{tab:BSSELOCAL}.

These results are even more significant from the perspective of complexity. Since each Random Forest regression result is based on an ensemble of regression-decision-trees with basically unconstrained complexity. While our RMA-based decision-making is based on only two major and one minor covariate features.

\begin{table}[h!]
\caption{Sum of squared error (SSE)}
\centering
\begin{tabular}{l|cc|cc|}
\cline{2-5}
  & \multicolumn{2}{c|}{Ours} & \multicolumn{2}{c|}{RF} \\\hline
\multicolumn{1}{|l|}{Patch}   &pfx\_x &pfx\_z &pfx\_x& pfx\_z\\\hline
\multicolumn{1}{|l|}{A1} &  0.00062 &  0.00061 & 0.00072&  0.00110\\
\multicolumn{1}{|l|}{A2}  & 0.00093 &  0.00266 & 0.00088 & 0.00149\\
\multicolumn{1}{|l|}{A3}  & 0.00016 &  0.00192 & 0.00080 & 0.00246\\
\multicolumn{1}{|l|}{B1}  & 0.00098 &  0.00271  &0.01796 & 0.00410\\
\multicolumn{1}{|l|}{B2}  & 0.00188 &  0.00560 & 0.00138&  0.00476\\
\multicolumn{1}{|l|}{B3}  & 0.00182  & 0.00375 & 0.00087 & 0.00546\\
\multicolumn{1}{|l|}{C1} & 0.00103  & 0.00306 & 0.00212 & 0.00197\\
\multicolumn{1}{|l|}{C2}   &0.00090 &  0.00296 & 0.00053 & 0.00677\\
\multicolumn{1}{|l|}{C3}  & 0.00136 &  0.00283&  0.00050 & 0.00241\\\hline
\end{tabular}%

\label{tab:SSEINV}
\end{table}

\begin{table}[h!]
\caption{Bivariate correlated sum of squared errors (covariance matrix $\Sigma$ computed based on all training nodes)}
\centering
\begin{tabular}{|l|cc|}\hline
 Patch & Ours & RF\\\hline
A1        &                       0.024   &                           0.599\\
A2         &                     0.061    &                          0.071\\
A3         &                      0.012   &                           0.467\\
B1         &                      0.034   &                           0.639\\
B2         &                     0.940    &                          1.230\\
B3         &                      0.078   &                           5.884\\
C1         &                     0.053     &                         0.005\\
C2         &                      0.079     &                         8.244\\
C3         &                     0.024    &                          0.307\\\hline
\end{tabular}%
\label{tab:BSSEALL}
\end{table}

\begin{table}[h!]
\caption{Bivariate correlated sum of squared errors (9 locality-specific covariance matrixes ${\Sigma_j}^(_{j=1}$ computed based on only training within each locality)}
\centering
\begin{tabular}{|l|cc|}\hline
 Patch & Ours & RF\\\hline
A1& 14.852& 123.574\\
A2&  6.657  &11.377\\
A3& 2.366 & 35.105\\
B1& 13.855 & 77.308\\
B2& 70.931  &58.974\\
B3& 15.402 &305.538\\
C1&  7.416  & 0.456\\
C2& 11.795 &332.379\\
C3&  3.333 & 23.411\\\hline
\end{tabular}%
\label{tab:BSSELOCAL}
\end{table}

\subsection{Manifold on \{``x0'', ``start\textunderscore speed'', ``end\textunderscore speed''\}}
The response feature ``end\textunderscore speed'' seemingly involves the simplest mechanism among the 7 response features. Its functional relationship with ``start\textunderscore speed'' or ``vx0'' (the releasing speed) is naturally expected to be linear. So feature ``start\textunderscore speed'' is an obvious major feature toward ``end\textunderscore speed''. Are there other major features? Not ``spin\textunderscore dir'' or ``spin\textunderscore rate''. Indeed we haven't found any one, other than the feature ``x0''. Different ranges of ``x0'' values seemingly lead to different pitcher-specific scatter plots of ``start\textunderscore speed'' against ``end\textunderscore speed''. It is indeed more-or-less so for all these pitch-types, as shown in Panels-C of Fig. \ref{fig:infogeoS}, and panels-B of Fig. \ref{fig:3Dxydir} respectively for slider and fastball. Further, it is noted that the feature ``x0'' is highly associative with pitcher-ID, as discussed in MCC section. It is natural to see that pitcher-ID is a minor feature, given ``x0'' is taken as major feature.

Upon the manifold of \{``x0'', ``start\textunderscore speed'', ``end\textunderscore speed''\}, we explore pitcher-specific relations of \{``x0'', ``start\textunderscore speed''\} to response feature ``end\textunderscore speed'' by considering the simple linear regression model fitting to each individual pitcher's data is:
\[
Y_i=\alpha + \beta_1 X_{1,i} + \beta_2 X_{2,i}+ \varepsilon_i
\]
with $Y_i$, $X_{1,i}$ and $X_{2,i}$ for $i-$th pitch's ``end\textunderscore speed'', ``x0'' and ``start\textunderscore speed'', respectively.

All pitcher-specific estimates of $\beta_2$, the slope value of ``start\textunderscore speed'' as shown in Table ~\ref{tab:sp-fast}. These statistically significant estimates collectively have a range $[0.783, 1.099]$ of values, which is much wider than slider and curveball's narrow ranges centering around $1.0$. This is striking because it is somehow against our intuition. The reasons underlying these pitchers' exceptional low values of $\beta_2$ needs requires further explanations and explorations. On the other hand, this manifold on \{``x0'', ``start\textunderscore speed'', ``end\textunderscore speed''\} gives rise clear message of inferential importance of major features as well as the minor ones. This involvement of pitcher-ID make the response manifold analytics (RMA) connected to MCC. That is, for prediction purpose, we need to proceed first to identify candidate pitcher-IDs, and then apply the linear regression inference. This is exactly in accord with our RMA-based inferential protocol.

\begin{table}[h!]
\caption{Fastball}
\centering
\begin{tabular}{lllllll}
Pitcher & intercept & x0      & start\_speed & residual std error & df   &  \\\hline
572971  & 0.038*    & -0.046* & 1.009*       & 0.027              & 1491 &  \\
457918  & 0.124*    & -0.028  & 0.785*       & 0.037              & 1759 &  \\
606160  & 0.012     & -0.136* & 0.981*       & 0.042              & 1242 &  \\
519242  & 0.025     & -0.039  & 0.928*       & 0.038              & 1649 &  \\
450203  & 0.067*    & -0.279* & 0.934*       & 0.033              & 1627 &  \\
572020  & 0.247*    & -0.220* & 0.870*       & 0.036              & 1713 &  \\
502748  & 0.085*    & -0.011  & 0.801*       & 0.032              & 1028 &  \\
448179  & 0.021     & -0.037  & 0.916*       & 0.036              & 1288 &  \\
435221  & 0.146*    & -0.045  & 0.783*       & 0.031              & 692  &  \\
448802  & 0.072*    & -0.011  & 0.861*       & 0.037              & 1470 &  \\
434378  & 0.018     & -0.112* & 0.931*       & 0.044              & 1988 &  \\
472610  & -0.024*   & 0.382*  & 0.952*       & 0.035              & 698  &  \\
594742  & 0.065*    & -0.005  & 0.825*       & 0.031              & 746  &  \\
622491  & -0.006    & -0.172* & 0.982*       & 0.039              & 922  &
\end{tabular}%
\label{tab:sp-fast}
\end{table}

\section{Conclusions from MCC and RMA perspectives.}
Via CEDA computing from MCC and RMA perspectives, we are able to provide information content fully reflecting discovered geometries and manifolds for three relatively small data sets taken from PITCHf/x database. By extracting and presenting collectives of multiscale geometric pattern categories as information content, we are able to see the system's key mechanisms and understand how they work on a realistic and visible platform. From these aspects, we advance current knowledge and intelligence. We believe that CEDA is capable of leading us to a fuller version of pitching dynamics based on data of an entire MLB season. As byproducts, we are able to point out fundamental misconceptions and dissect uncertainty of popular machine learning methodologies. Such products imply and confirm high potential merits of joint endeavors of our CEDA and many other machine learning approaches. Such joint computing endeavors likely help researchers exploring in many complex systems in all sciences, businesses and industries because of its universal applicability on all data types. Though, information content resulted from CEDA does suit for a computer or a robot very well, it is by design very readable, visible and explainable for human. Overall our CEDA computing theme could become the necessary computing paradigm for extracting full information content from all databases to sustain Explainable Artificial Intelligence (E.A.I.).

Our (CEDA) computing theme on a MCC setting makes good use of a chain of asymmetry of a chain of mixing geometries pertaining to a chain complementary feature-sets. The resultant MCC information content is presented in an ordered series of tables of multiscale details of mixing geometric pattern categories. It is also critical to note that a selected chain means a way of synthesizing results from different feature-sets. Such discoveries and synthesis reiterate the essence of CEDA for discovering diverse perspectives of heterogeneity within any data set from a complex system. This is a true basis of knowledge and intelligence under any MCC setting. 		 
Our stochastic partial ordering and its resultant dominance matrix for computing a label embedding tree (LET) is very simple in ideas and efficient in computing cost. Consequently, LET is very robust with respect to outlier and overlapping edges of point-clouds. Hence LET is a reliable geometry on a label space to facilitate the developments of serial binary competitions with length of order $O(\log_2 K)$. This is why the complexity of MCC information content is rather low.

From RMA perspective, our RMA-based inferential protocol is fundamentally simple and efficient.  It does particularly suit for predictive decision-making on multiple response variables. Another critical importance RMA is the fact that it brings out geometric information content of coupling multiple response with multiple covariate features without any man-made structural assumptions. So, our RMA is a natural choice under any setting with apparent nonlinearity.

Further, we would like to emphasize that our RMA-based inferential protocol via CEDA theme computing is indeed a universal methodology for studying a complex system from data science standpoint. This universality rests on categorical data structure. This is a surprising reality. This reality starts from a simple fact: each feature of any data type: continuous, discrete and categorical, has a possibly-gapped histogram \cite{hsiehroy}. Thus, $m$ response features define an ensemble of hypercubes framed by bins of histograms of involving features. Their intertwined associations among these $m$ response features are fibers constituting a response ``manifold'' in a form of $m$-D contingency hyperspace with much reduced complexity. This reduced complexity is meant to be reflecting that majority of hypercubes framed by the $m$ histograms are empty. That is, the ensemble of occupied hypercubes only retains a tiny proportion.

The above idea of response manifold is further extended into a manifold of a targeted set of $m$ response features coupled with a set of $k$ major covariate features.  The selection of candidate major covariate features is highly regulated by an existing road map: the mutual conditional entropy matrix of all features, which are also resting on their categorical structures. Upon such a $m+k$-D manifold, we have natural and well-defined localities. We then can check and discover minor features via Shannon entropy. This construction of a $m+k$-D manifold for RMA is indeed universal in the sense that it works for all data types.

Therefore, we can ``see'' and ``read'' geometric information on the global and local scales, and at the same time visualize how each minor feature's heterogeneity acting on localities across the entire manifold. Our predictive decision-making approach is just a natural way of ``looking'' into geometric pattern information.

\section{References}


\bibliographystyle{unsrt}

\end{document}